\documentclass[longauth]{aa}  

\usepackage{color}
\usepackage{graphicx}
\usepackage{natbib}
\usepackage{enumitem}
\usepackage{txfonts}
\usepackage{multirow}
\usepackage{rotating}
\usepackage{longtable}
\titlerunning{XXL spectrophotometric sample and mass functions}
\authorrunning{V. Guglielmo et al.}
\begin{document}

\title{The XXL Survey: XXII. The XXL-North spectrophotometric sample and galaxy stellar mass function in X-ray detected groups and clusters}

\subtitle{}

\author{V. Guglielmo \inst{1,} \inst{2,}\inst{3} \and B.M. Poggianti \inst{1} \and B. Vulcani \inst{4,}\inst{1} \and C. Adami \inst{2} \and F. Gastaldello \inst{5} \and S. Ettori \inst{6} \and S. Fotoupoulou \inst{7} \and E. Koulouridis \inst{10} \and M. E. Ramos Ceja \inst{8} \and P. Giles \inst{9} \and S. McGee \inst{29} \and{B. Altieri} \inst{35} \and I. Baldry \inst{11} \and M. Birkinshaw \inst{9} \and{M. Bolzonella} \inst{6} \and A. Bongiorno \inst{12} \and M. Brown \inst{13} \and L. Chiappetti \inst{5} \and S. Driver \inst{14,}\inst{34} \and A. Elyiv \inst{15,}\inst{16} \and A. Evrard \inst{17} \and B. Garilli \inst{5} \and M. Grootes \inst{18} \and L. Guennou \inst{19} \and A. Hopkins \inst{20} \and C. Horellou \inst{21} \and A. Iovino \inst{22} \and C. Lidman \inst{20} \and J. Liske \inst{23} \and S. Maurogordato \inst{24} \and M. Owers \inst{25} \and F. Pacaud \inst{8} \and S. Paltani \inst{7} \and M. Pierre \inst{10} \and M. Plionis \inst{26,}\inst{27,}\inst{28} \and T. Ponman \inst{29} \and A. Robotham  \inst{14} \and T. Sadibekova \inst{10,}\inst{30} \and M. Scodeggio \inst{5} \and M. Sereno \inst{6} \inst{15} \and V. Smol\v ci\'c \inst{31} \and R. Tuffs \inst{18} \and I. Valtchanov \inst{35} \and C. Vignali \inst{6,}\inst{15} \and J. Willis \inst{33}}

\institute{INAF- Osservatorio astronomico di Padova, Vicolo Osservatorio 5, IT-35122 Padova, Italy\\
              \email{valentina.guglielmo@oapd.inaf.it}
              \and
             Aix Marseille Universit\'e, CNRS, LAM (Laboratoire d'Astrophysique de Marseille) UMR 7326, F-13388, Marseille, France
             \and Department of Physics and Astronomy, University of Padova, Vicolo Osservatorio 3, IT-35122 Padova, Italy
            \and School of Physics, University of Melbourne, VIC 3010, Australia
            \and Istituto di Astrofisica Spaziale e Fisica Cosmica Milano, Via Bassini 15, IT-20133 Milan, Italy
            \and INAF, Osservatorio Astronomico di Bologna, via Ranzani 1, IT-40127 Bologna, Italy
            \and Department of Astronomy, University of Geneva, ch. d’Ecogia 16, CH-1290 Versoix, Switzerland
            \and Argelander Institut für Astronomie, Universität Bonn, Auf dem Huegel 71, DE-53121 Bonn, Germany
            \and School of Physics, HH Wills Physics Laboratory, Tyndall Avenue, Bristol, GB-BS8 1TL, UK
            \and IRFU, CEA, Universit\'e Paris-Saclay, F-91191 Gif-sur-Yvette, France\\
Universit\'e Paris Diderot, AIM, Sorbonne Paris Cit\'e, CEA, CNRS, F-91191
Gif-sur-Yvette, France
            \and Astrophysics Research Institute, Liverpool John Moores University, IC2, Liverpool Science Park, 146 Brownlow Hill, Liverpool GB-L3 5RF, UK
            \and Osservatorio Astronomico di Roma (INAF), via Frascati 33, IT-00078 Monte Porzio Catone (Rome), Italy
            \and School of Physics and Astronomy, Monash University, Clayton, Victoria AU-3800, Australia 
            \and International Centre for Radio Astronomy Research (ICRAR), The University of Western Australia, M468, 35 Stirling Highway, Crawley, WA AU-6009, Australia
            \and Dipartimento di Fisica e Astronomia, Alma Mater Studiorum – Università di Bologna, viale Berti Pichat 6/2, IT-40127 Bologna, Italy
            \and Main Astronomical Observatory, Academy of Sciences of Ukraine, 27 Akademika Zabolotnoho St., UA-03680 Kyiv, Ukraine
            \and Department of Physics and Michigan Center for Theoretical Physics, University of Michigan, Ann Arbor, MI Us-48109, USA
            \and ESA/ESTEC, Noordwijk 2201 AZ, The Netherlands
            \and Astrophysics and Cosmology Research Unit, University of KwaZulu-Natal, ZA-4041 Durban, South Africa
            \and Australian Astronomical Observatory, PO BOX 915, North Ryde, AU-1670, Australia
            \and Chalmers University of Technology Onsala Space Observatory, SE-439 92 Onsala, Sweden
            \and INAF-Osservatorio Astronomico di Brera, via Brera, 28, IT-20159 Milano, Italy
            \and Universit\"{a}t Hamburg, Hamburger Sternwarte, Gojenbergsweg 112, DE-21029 Hamburg, Germany
            \and Laboratoire Lagrange, UMR 7293, Universit\'e de Nice Sophia Antipolis, CNRS, Observatoire de la C\^{o}te d’Azur, FR-06304 Nice, France
            \and Department of Physics and Astronomy, Macquarie University, NSW 2109, Australia and Australian Astronomical Observatory PO Box 915, North Ryde NSW AU-1670, Australia
            \and Aristotle University of Thessaloniki, Physics Department, GR-54124 Thessaloniki, Greece
            \and Instituto Nacional de Astrof\'{i}sica \'{O}ptica y Electr\'{o}nica, AP 51 y 216, 72000 Puebla, MX-Mexico
            \and IAASARS, National Observatory of Athens, GR-15236 Penteli, Greece
            \and School of Physics and Astronomy, University of Birmingham, Edgbaston, Birmingham GB-B15 2TT,UK
            \and Ulugh Beg Astronomical Institute of Uzbekistan Academy of Science, 33 Astronomicheskaya str., Tashkent, UZ-100052, Uzbekistan
            \and Department of Physics, University of Zagreb, Bijenicka cesta 32, HR-10000 Zagreb, Croatia
            \and Space sciences, Technologies and Astrophysics Research (STAR) Institute, Universit\'e de Li\`ege, 19c All\'ee du Six Ao\^{u}t, BE-4000 Li\`ege, Belgium
            \and Department of Physics and Astronomy, University of Victoria, 3800 Finnerty Road, Victoria, BC, Canada           
            \and SUPA, School of Physics \& Astronomy, University of St Andrews, North Haugh, St Andrews GB-KY16 9SS, UK
            \and Herschel Science Centre, European Space Astronomy Centre, ESA, S-28691 Villanueva de la Ca\~{n}ada, Spain}

\date{Received xxx; accepted yyy}

\abstract
% 5 {} token are mandatory
% context heading (optional)
{The fraction of galaxies bound in groups in the nearby Universe is high (50\% at $z \sim 0$). Systematic studies of galaxy properties in groups are important in order to improve our understanding of the evolution of galaxies and of the physical phenomena occurring within this environment.}
% aims heading (mandatory)
{We have built a complete spectrophotometric sample of galaxies within X-ray detected, optically  spectroscopically confirmed groups and clusters (G\&C), covering a wide range of halo masses at $z\leq 0.6$.}%low and intermediate redshift.}
% methods heading (mandatory)
{In the context of the XXL survey, we analyse a sample of 164 G\&C in the XXL-North region (XXL-N), at $z \leq 0.6$, with a wide range of virial masses ($\rm 1.24 \times 10^{13} \leq M_{500} (M_\sun) \leq 6.63 \times 10^{14}$) and X-ray luminosities ($\rm 2.27 \times 10^{41} \leq L^{XXL}_{500} (erg \, s^{-1}) \leq 2.15 \times 10^{44}$). The G\&C are X-ray selected and spectroscopically confirmed. We describe the membership assignment and the spectroscopic completeness analysis, and compute stellar masses. As a first scientific exploitation of the sample, we study the dependence of the galaxy stellar mass function (GSMF) on global environment.}
% results heading (mandatory)
{We present a spectrophotometric characterisation of the G\&C and their galaxies. The final sample contains 132 G\&C, 22111 field galaxies and 2225 G\&C galaxies with $r$-band magnitude $<20$. Of the G\&C, 95\% have at least three spectroscopic members, and 70\% at least ten. The shape of the GSMF seems not to depend on environment (field versus G\&C) or X-ray luminosity ( used as a proxy for the virial mass of the system). These results are confirmed by the study of the correlation between mean stellar mass of G\&C members and $L^{XXL}_{500}$.We release the spectrophotometric catalogue of galaxies with all the quantities computed in this work.}
% conclusions heading (optional)
{As a first homogeneous census of galaxies within X-ray spectroscopically confirmed G\&C at these redshifts, this sample will allow environmental studies of the evolution of galaxy properties.}


\maketitle
%
%-------------------------------------------------------------------

\section{Introduction}
According to the commonly adopted lambda cold dark matter ($\Lambda$CDM) cosmological model,  structures grow in a  hierarchical  fashion: as time proceeds, smaller structures merge to form  larger ones. This process implies that the fraction of galaxies located in groups  increases with cosmic time, up to Local Universe values (\citealt{Huchra1982,Tully1987,Eke2004a,Berlind2006,Knobel2009}). Today, over 50\% of galaxies are in  groups, span a wide range in local density, and show properties that range from cluster-like to field-like (\citealt{Zabludoff1998}). Therefore groups are a key environment to investigate galaxy evolution and to provide a clear framework to study the nature of the physical mechanisms that lead to galaxy transformation.

The physical processes operating in groups are poorly understood. For example, to what extent do gravitational interactions and the intra-group medium determine the morphology and star formation properties of galaxies residing in and around groups? 
It has been proposed that galaxy-group interactions like halo gas stripping (`strangulation') can remove warm and hot gas from a galaxy halo, efficiently cutting off the  gas supply for star formation (\citealt{Larson1980,Cole2000,Balogh2000,Kawata2008}). Alternatively, mergers/collisions and close tidal encounters among group member galaxies can strongly alter the galaxy structure \citep{Toomre1972} and may result in star formation quenching. In addition, at the typical velocity dispersions of bound groups, galaxy-galaxy harassment (i.e. frequent  galaxy encounters) leads to the morphological transformation of disc galaxies. Indeed, it results in the loss of a galaxy's gaseous component, partly ablated and partly falling into the centre, entailing a dramatic conversion of discs into spheroidals. The combination of tides and ram pressure stripping efficiently removes the gas from spirals, quenching the star formation in galactic discs, while triggering it in the  arms and at the leading edge of gaseous disc, with the net result of a suppression of star formation on timescales of several Gyr \citep{Moore1996}. Both  high-resolution N-body simulations and  semi-analytic models of galaxy evolution have shown that these processes play a very important role in the formation of galaxy populations \citep{Barnes1996,Springel2001,Mihos2004,Kang2005,Murante2007,Wang2007,Cox2008,Font2008,Somerville2008,Weinmann2010,Guo2011,Henriques2015}.

Even though galaxy groups are  more common than more massive structures, they are much more difficult to detect because of their lower density contrast with respect to the background galaxy population. 
 
Until recently, the difficulties in obtaining large, unbiased samples of groups have forced most studies to use small samples selected, for example, from the Hickson compact group catalogue \citep{Hickson1989}, from the CfA redshift survey \citep{Geller1983,Moore1993}, and from X-ray surveys \citep{Henry1995,Mulchaey2003}. Only with the advent of large galaxy redshift surveys, such as the Two Degree Field Galaxy Redshift Survey (2dFGRS), the Sloan Digital Sky Survey (SDSS) and the Canadian Network for Observational Cosmology Redshift Survey (CNOC2),  has it become possible to generate large group catalogues in the local Universe (e.g. \citealt{Huchra1982,Ramella1989,Ramella1997,Hashimoto1998,Ramella1999,Tucker2000,Martinez2002,Balogh2004,Eke2004b,Calvi2011,Moustakas2013,Miniati2016}) and at intermediate redshift \citep{Carlberg2001,Wilman2005,Giodini2012,George2013}.

Overall, while  many of the observational studies so far have focused on large surveys at low redshift,  the  common group-scale environments and their evolution over time are still poorly known. 
At $z<1$, COSMOS \citep{Scoville2007} is one of the most studied fields. Several teams have assembled a number of group catalogues (e.g. \citealt{Knobel2009,Giodini2012,George2013}) outlining some trends. 
For instance, \cite{Presotto2012} found that  galaxies with $\log(M_\ast/M_\sun) \geq 10.6$ do not display any strong  dependence of the fractions of red/blue objects on groupcentric distance, while for galaxies with $9.8 \leq \log(M_\ast/M_\sun) \leq 10.6$ there is a radial dependence in the changing mix of red and blue galaxies. This dependence is most evident in poor groups, whereas richer groups do not display any obvious trend of the blue fraction. Mass segregation shows the opposite behaviour: it is visible only in rich groups, while poorer groups have a a constant mix of galaxy stellar masses as a function of radius. \cite{George2013} found a decline in low-mass star-forming and disc-dominated galaxies from field to groups. This behaviour is accompanied by an increase in the quenched fraction of intermediate-type galaxies (disc+bulge) from field to groups, while bulge-dominated systems show only weak evolution. \cite{Giodini2012} found significant differences in the build-up of the quenched population from field to group galaxies at low stellar masses, while no differences are found for star-forming galaxies.

Among the many galaxy properties that can be studied, the galaxy stellar mass function (GSMF) is an important diagnostic tool for performing a census of galaxy properties, and provides a powerful means of comparison between the populations of galaxies in different environments. In particular, its shape  and its evolution provide  important insights into the processes that contribute to the growth in stellar mass of galaxies with time and that drive the formation and evolution of galaxies in different environments.

The GSMF has been extensively studied in deep fields for galaxies of different colours and morphological types \citep{Bundy2006,Baldry2008,Pozzetti2010,Vulcani2011} and in different environments \citep[e.g.][]{Balogh2001,Yang2009,Calvi2013,Vulcani2011,Vulcani2012,Vulcani2013,Vulcani2014,Davidzon2016,Muzzin2013,VanderBurg15,Nantais2016}. Its shape has been described by a Schechter or a double Schechter function \citep{Schechter1976}. When fitted to the data, the shape of this function changes both as a function of the galaxy type (star-forming/passive, or morphological type) and of the environment.

Many different parametrisations of the environment can be adopted. When considering galaxies belonging to a structure, both in the local Universe and at higher redshift, it has been shown that the shape of the GSMF shows very little variation from isolated systems to massive clusters (e.g. \citealt{Calvi2013,Vulcani2013,VanderBurg15,Nantais2016}, but see \citealt{Yang2009}). In contrast, when considering local density estimates, the GSMF seems to depend on environment, being steeper in less dense environments \citep[e.g.][]{Baldry2006,Bolzonella2010, Vulcani2012,Davidzon2016,Etherington2016}.

In this paper we assemble a catalogue of galaxies in X-ray selected groups and clusters (G\&C) from the XXL Survey in the redshift range $0<z<1.5$, and pay particular attention to galaxies at $z\leq 0.6$.
The XXL Survey \citep[hereafter XXL Paper I]{Pierre2016}, is an extension of the XMM-LSS 11 $\rm deg^2$ survey \citep{Pierre2004}, and is made up of 622 XMM pointings covering a total area of $\sim 50 \, {\rm deg^2}$ and reaching a sensitivity of $\rm \sim 5 \times 10^{-15} erg \, s^{-1} \, cm^{-2}$ in the [0.5-2] keV band for point sources.
With respect to previous G\&C catalogues at similar redshifts, the sample covers a much wider area in the sky, with the advantage of diminishing the cosmic variance, and includes G\&C confirmed spectroscopically, which span a wide range in X-ray luminosity ($\rm 2.27 \times 10^{41} \leq L^{XXL}_{500} (erg \, sec^{-1}) \leq 3.5 \times 10^{44}$) and therefore virial masses ($\rm 8.72 \times 10^{12} \leq M_{500} (M_\sun) \leq 6.64 \times 10^{14}$). The G\&C membership determinations are  robust, being based on spectroscopic redshifts and on virial masses derived from X-ray quantities via scaling relations (Adami et al. in prep., hereafter XXL Paper XX).

As a first exploitation of the catalogue, we investigate the behaviour of the GSMF in the redshift range $0<z\leq 0.6$ as a function of global environment (G\&C versus field) and as a function of X-ray luminosity.
The advantage of this work  is  that it  is based on a large, homogeneous X-ray selected sample of G\&C that are
spectroscopically confirmed and span a wide range in X-ray luminosity, therefore uniformly probing a wide range of halo masses. 

The paper is organised as follows. Section \ref{datasamples} presents the data sample and the photometric and spectroscopic catalogues, along with the spectroscopic completeness. Section \ref{membership} characterises the environments in which galaxies are embedded, and Section \ref{masses} the mass estimates. 
Section \ref{sp_cat} presents the catalogue that we publicly release. Section \ref{results} shows the results of our analysis of the galaxy stellar mass function, while Section \ref{conclusions} summarises our work.

Throughout the paper, we assume $\rm H_0 = 69.3 km \, s^{-1} \,Mpc^{-1}, \, \Omega_M = 0.29, \, \Omega_{\Lambda} = 0.71$.  We adopt a \cite{Chabrier2003} initial mass function (IMF) in the mass range 0.1-100 $\textrm{M}_{\odot}$. 

%--------------------------------------------------------------------
\section{Data sample}
\label{datasamples}
This study is based on X-ray selected G\&C, drawn from the sample of structures % at intermediate redshifts 
identified within the XXL survey. % (Paper I).

In this section we describe the XXL X-ray observations and the final database and catalogues that are used in this work. While our scientific analysis will be based only on the XXL North field (XXL-N), for the sake of completeness in the following we also report on the data for the XXL South field (XXL-S). At the time of writing, the galaxy spectroscopic coverage of the latter field is insufficient to have much statistical weight.
%For the sake of simplicity, from now on we will define as ``structures'' all the X-ray extended sources, regardless of their size. In fact, as we will see later in this section, our sample spans a wide range in virial masses and radii of the structures, including both smaller (i.e. groups) and larger structures (i.e. clusters).

\subsection{X-ray observations and   database of the G\&C}

The description of the practical requirements and of the observing strategies which prevailed in the definition of the XXL X-ray sample are fully described in Paper I.
The final selected areas were (1) the North region: the XMM-LSS field, with 10 ks observations over 25 $\rm deg^2$ in the CFHTLS-W1 Field (2h23 -05d00) with 11 deg$^2$ previously covered with XMM exposures of 10-20 ks \citep{Pierre2004} plus the XMM-Subaru Deep Survey \citep{Ueda2008} and  (2) the South region: the BCS/XMM field with the same 10 ks exposure time as the north, covering another area of 25 $\rm deg^2$ (23h00 -55d00).
The flux limit for 10ks observations is $ 4 \times 10^{-15}$ and $2 \times 10^{-14} {\rm erg \, cm^{-2} \, sec^{-1}}$ in the soft ([0.5-2] keV) and hard ([2-10] keV) bands, respectively.

The data processing of X-ray observations and the sample selection are described in detail in \citet[hereafter XXL Paper II]{Pacaud2016}.
Briefly, data were processed with the \textsc{Xamin} v3.3.2 pipeline for the detection and classification of X-ray faint extended sources, a dedicated pipeline already used in the pilot XMM-LSS project \citep{Pacaud2006,Clerc2012} to generate and process images, exposure maps and detection maps.
The procedure is based on two parameters named \verb!ext! and \verb!ext_stat!, which are both functions of the structure apparent size, flux and local XMM sensitivity: a detection enters the extended candidate list when it has an \verb!ext! greater than 5$^{\prime\prime}$ and a likelihood \verb!ext_stat! greater than 15. Extensive simulations enabled the creation of different classes for structures on the basis of the level of contamination from point-sources:
\begin{itemize}
\item class 1 (c1) includes the highest surface brightness extended sources, which have an \verb!ext_stat! $>33$, detection statistic \verb!ext_det_stat! $>32$ and are identified such that no point sources are misclassified as extended;
\item class 2 (c2) includes sources with $15 < \verb!ext_stat! < 33$  showing a 50\% contamination rate. c2 G\&C are fainter than those in c1. Contaminating sources include saturated point sources, unresolved pairs, and sources strongly masked by CCD gaps, for which not enough photons were available to permit reliable source characterisation; 
\item class 3 (c3) class includes sources at the survey sensitivity limit, and so is likely to contain G\&C at high redshift. c3 G\&C are faint objects and therefore have less well-characterised X-ray properties.
\end{itemize}

The list of c1, c2, c3 detections are hosted in the Saclay database \footnote{http://xmm-lss.in2p3.fr:8080/xxldb/} (administered by Jean Paul Le F\`evre), which contains 455 analysable extended sources: 207 ($\sim 46\%$) of them are classified as c1 sources, 194 ($\sim 43\%$) are c2 sources, and the remaining 51 ($\sim 11\%$) are c3 sources. 
%A subsample of 100 G\&C was selected from the entire sample (all classified as C1 and C2), containing the highest signal-to-noise objects, and was the subject of many of the first series of XXL publications. \textbf{Among them, those relevant in this work are XXL Paper II, \citet[hereafter XXL Paper III]{Giles2016}, \citet[hereafter XXL Paper IV]{2016A&A...592A...4L}}}

Among the 455 XXL G\&C, 264 are in the XXL-N area.

The spectroscopic confirmations of the nature of the candidate G\&C and of their redshifts were performed using an iterative semi-automatic process, very similar to the one already used for the XMM-LSS survey (e.g. \citealt{Adami2011}). The procedure is described in detail in  XXL Paper XX, and can be summarised as follows: 

\begin{itemize}
\item Within the X-ray contours, the available spectroscopic redshifts from the XXL spectroscopic database (see Sec. \ref{spec_data}) were selected;
\item These redshifts were sorted by ascending order to identify significant gaps ($\Delta z>$0.003) in their distribution;
\item If one or more concentrations in both physical and redshift space appeared (more than three galaxies), the aggregate of galaxies closer to the X-ray centre or that including the Brightest Cluster Galaxy (BCG) were selected as `group population'. For the vast majority of the cases, a single concentration emerged (see XXL Paper XX for a more precise discussion on multiple systems);
\item If no concentration appeared, a single galaxy with measured redshift which was likely to be a BCG was selected. This did not exclude superposition effects, but the probability of such a configuration is low;
\item If neither of the two previous criteria was satisfied, the candidate structure could not be confirmed. If one of the two previous criteria was satisfied, the median value of the redshift of the preliminary `G\&C population' was assumed to be the G\&C redshift. This allowed us to compute the angular radius of a 500 kpc (physical) circle;
\item The whole process was repeated with all available redshifts within a 500 kpc radius instead of those within the X-ray contours to obtain the final G\&C redshift.
\end{itemize}

\begin{figure}
\begin{center}
\includegraphics[scale=0.47,clip, trim=8 10 40 30]{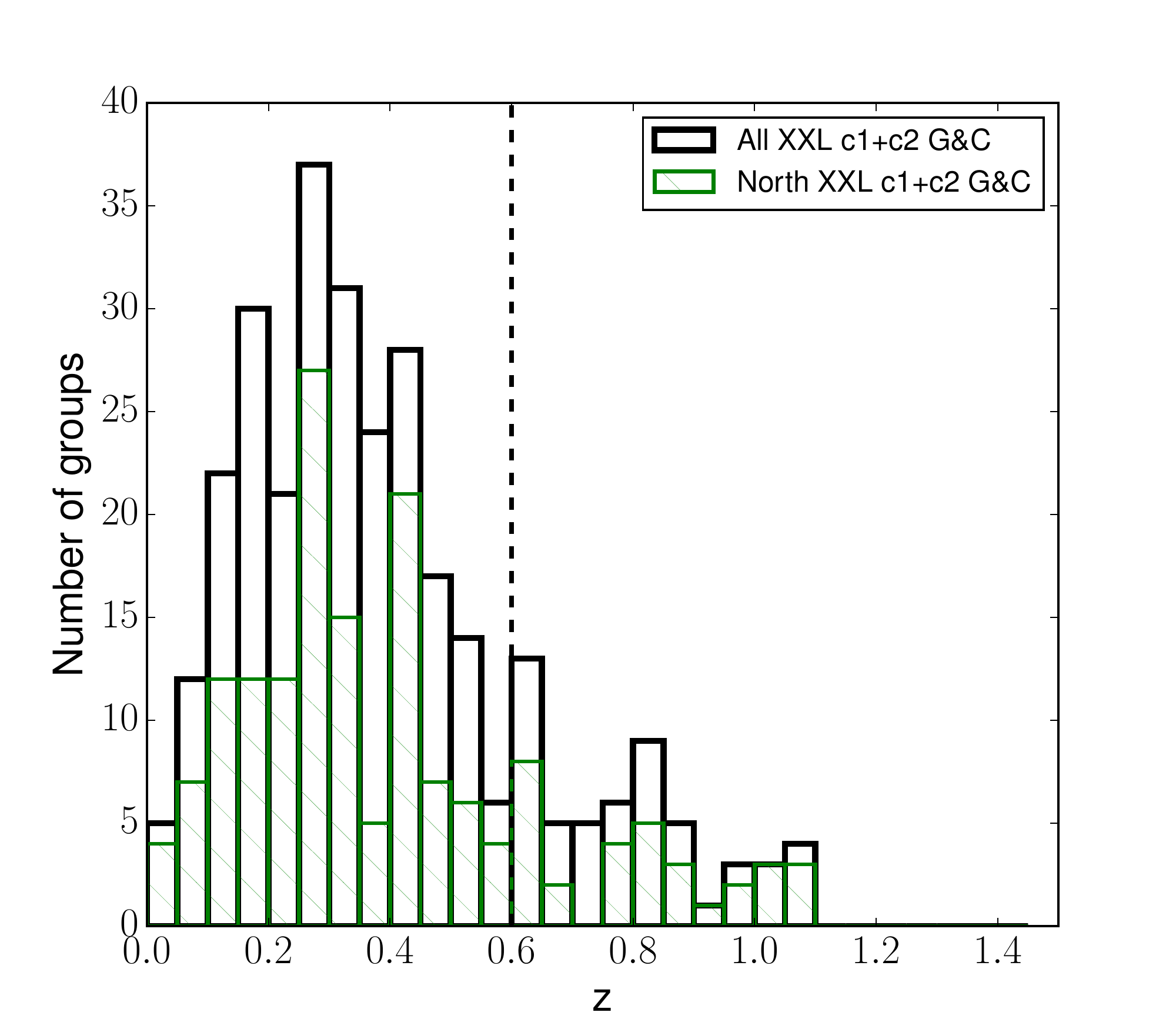}
\caption{ Redshift distribution of all 302 spectroscopically confirmed XXL c1+c2 G\&C (black), and of the 164 in the XXL-N area (green). The vertical black dashed line corresponds to $z=0.6$, the maximum redshift of G\&C considered in this work.
\label{z_histo}}
\end{center}
\end{figure}

This procedure identified 341 spectroscopically confirmed G\&C in the whole XXL sample, 202 of which in the XXL-N. Among the latter, 27 G\&C were confirmed considering only the BCG, 44 G\&C considering the BCG and another concordant galaxy.
The final fractions of c1, c2 and c3 G\&C in the whole (XXL-N) sample are 54\% (52\%), 35\% (30\%), and 11\% (18\%), respectively. 

Detailed information and global statistics about the XXL G\&C sample can be found in XXL Paper XX, which publishes 341 confirmed G\&C plus all c1 non-confirmed X-ray extended sources, for a total of 365 structures. The 222 c1+c2 G\&C (233 c1+c2+c3) with fluxes brighter than $\sim$1.3 $\times$10$^{-14}$  $\rm erg \, s^{-1} \, cm^{-2}$  underwent dedicated X-ray luminosity and temperature measurements. In order to have homogeneous estimates for the complete sample, we applied scaling relations based on the $r=300$ kpc count-rates (see XXL Paper XX). The  $L^{XXL}_{500}$, $L_{bol}$, $T$, $M_{500}$, and $r_{500}$\footnote{$r_{500}$ is defined as the radius of the sphere inside which the mean density is 500 times the critical density $\rho_c$ of the Universe at the cluster's redshift; $M_{500}$ is then by definition equal to $4/3 \pi 500 \rho_c r_{500}^3$} values used in the current paper are also extracted from XXL Paper XX, where a description of the scatter with respect to the direct measurements (when available) is also given.

\begin{figure}
\begin{center}
\includegraphics[scale=0.33,clip, trim=8 8 15 30]{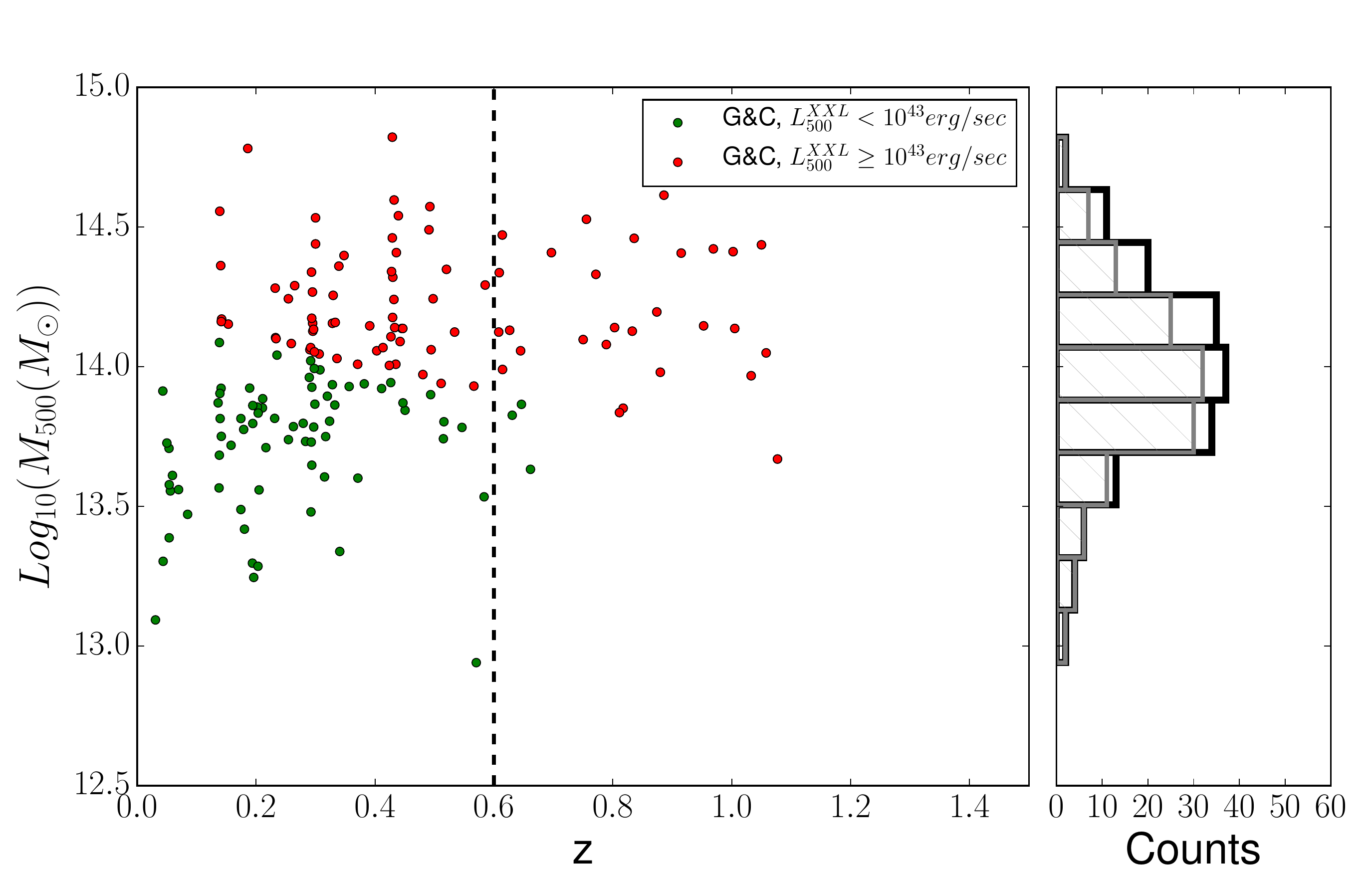}
\includegraphics[scale=0.33,clip, trim=8 8 15 30]{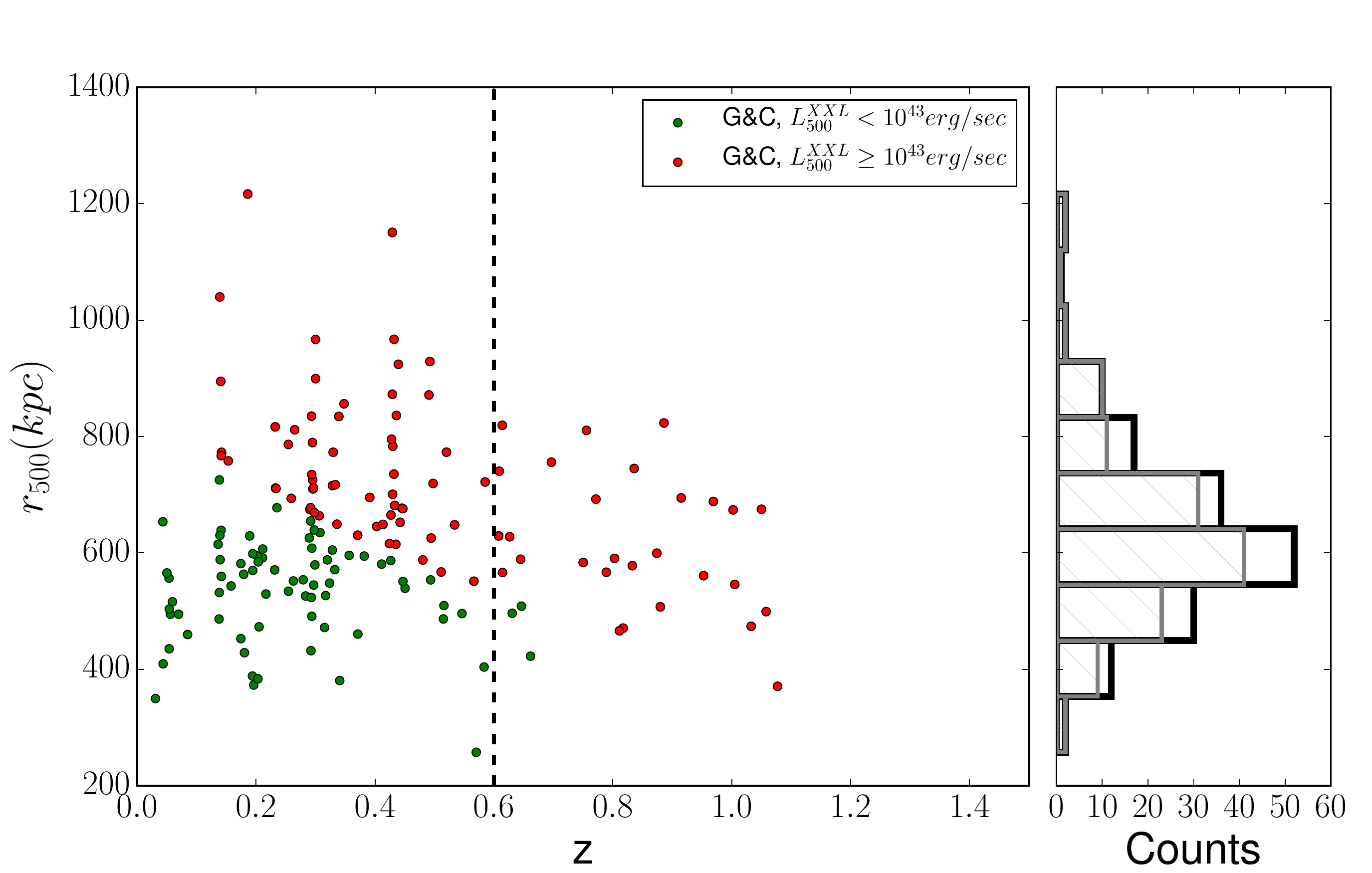}
\includegraphics[scale=0.305,clip, trim=8 0 15 30]{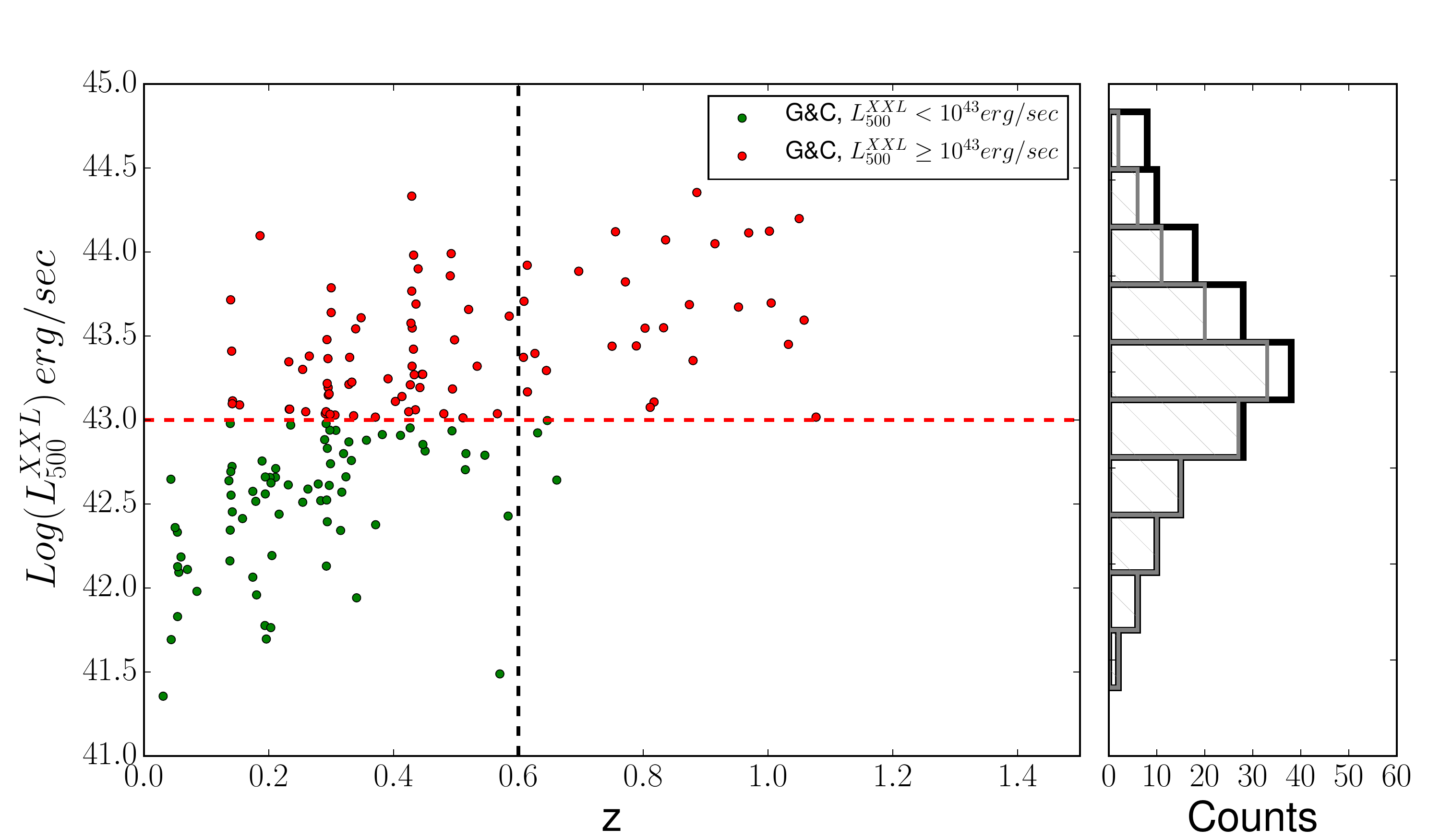}
\caption{Main panels: $M_{500}$ (upper), $r_{500}$ (middle), $L^{XXL}_{500}$ (bottom) vs. redshift for the 164  XXL-N c1+c2 G\&C with estimates of $M_{500}$ and  $r_{500}$. The distribution of the same quantities is shown in the corresponding right panels, where black histograms refer to all G\&C in the sample, and grey hatched histograms are for $z \leq 0.6$ G\&C. In the main panels, the vertical black dashed line corresponds to $z=0.6$, the maximum redshift of G\&C considered in this work. High-luminosity G\&C are marked in red, low-luminosity G\&C in green. In the bottom panel, the horizontal red dashed line corresponds to the luminosity  used to separate the G\&C into two classes (see text for details).}
\label{mrl500_z}
\end{center}
\end{figure}

Given the high uncertainties on X-ray properties derived for c3 G\&C, in the following we will consider only c1+c2 G\&C.
The redshift distribution of the c1+c2 G\&C sample is shown in Fig. \ref{z_histo}, where the histogram of the XXL-N sample is overlaid in green. A large fraction of X-ray G\&C are located at $\rm z \leq 1.0$, and in particular the median redshift of the sample is $z=0.339$ ($z=0.335$ when only the North field is considered).
Figure \ref{mrl500_z} shows how the $M_{500}$, $r_{500}$, and the temperature vary with redshift, for the 164 c1+c2. The G\&C found in the XXL-N field are, as already mentioned at the beginning of Sec. \ref{datasamples}, the main focus of this paper. The subpanels on the right show the distributions of the same quantities. The G\&C are divided into two classes according to their X-ray luminosity to study separately the properties of high- and low-luminosity G\&C. We use as a threshold the value $\rm L^{XXL}_{500} =\rm 10^{43} \, erg \, s^{-1}$, which corresponds approximately to the median value of the X-ray luminosity of the sample. 
Overall, selection effects emerge: at higher redshift the survey detects only the most massive G\&C.
The median $\rm M_{500}$ is $\rm (9.54 \pm 0.80) \times 10^{13} M_{\odot}$.
This indicates that roughly half of them should be properly qualified as clusters, since they have a mass $M_{500}\geq 10^{14}M_\sun$. The remaining half of them are more properly groups.
The distribution of $r_{500}$ resembles that of $M_{500}$, as expected given that these two quantities are closely related. 

The G\&C at $z \leq 0.6$ are used in the study of the galaxy stellar mass function (Sec. \ref{results}). In this redshift range, there is a wide range of virial masses ($\rm 8.72 \times 10^{12} \leq M_{500} (M_\sun) \leq 6.63 \times 10^{14}$) and of X-ray luminosities ($\rm 2.27 \times 10^{41} \leq L^{XXL}_{500} (erg \, s^{-1}) \leq 3.5 \times 10^{44}$).

\subsection{Photometric and photo-z databases}\label{photo}
We resort on different photometric observations that covered the XXL-N region. 
The largest contribution comes from  the CFHT Legacy Survey  \citep[CFHTLS]{2007AAS...210.6202V} and we rely on the Wide1 (W1) T0007 data release. 
Observations were obtained with  the 3.6m Canada-France-Hawaii Telescope (CFHT), using the MegaCam wide-field optical imaging facility. The MegaCam  camera consists of 36 CCDs of $2048 \times 4612$ pixels each and covers a field of view of 1 deg$^2$ with a resolution of 0.186 arcsec per pixel. The data cover the observed wavelength range 3500\AA $< \lambda < 9400$\AA $\,$ in the $u^*$, $g^\prime$, $r^\prime$, $i^\prime$, $z^\prime$ filters.  
We note that the MegaCam filter $i^\prime$ was broken during the survey and a new $i^\prime$ band filter was introduced (``$y^\prime$''). Both filters are considered and included separately in the catalogues.

W1 only covers the XXL region $30.17771 \leq RA (deg) \leq 38.8223$ and $-11.22814 \leq DEC (deg) \leq -3.70516$. 
To cover part of the remaining regions, we also exploit observations done by the MegaPipe Group GRZ programme \citep{2008PASP..120..212G} in the  $g$, $r$, $z$ bands. These observations cover the following  areas:
\begin{itemize}
\item Field A:  $\rm  35.10541 \leq RA \, (deg) \leq 36.09985$, \\ $\rm -3.78505 \leq DEC \, (deg) \leq -2.73612$,
\item Field B:  $\rm 36.06188 \leq RA \, (deg) \leq 37.05696$, \\ $\rm -3.78826 \leq DEC \, (deg) \leq -2.73855$.
\end{itemize}

For the W1 Field, we used the catalogue containing photometric redshifts computed from the Laboratoire d'Astrophysique de Marseille (LAM) in collaboration with Terapix using the spectral energy distribution (SED) fitting software  LePhare\footnote{www.lam.oamp.fr/arnouts/LEPHARE.html} \citep{Arnouts1999,Arnouts2002,Ilbert2006}.
The code consists of a set of Fortran programs and computes photometric redshifts with a standard $\chi^2$ method using SED fitting technique.
The Terapix\footnote{Traitement \'El\'ementaire R\'eduction et Analyse de PIXel (\cite{2007ASPC..376..507B}) is an astronomical data reduction centre dedicated to the processing of very large data flows from digital sky surveys (e.g. CFHTLS, WIRDS or WUDS, NGVS, CFHQSIR, KIDS/VIKING, UltraVISTA) and giant panoramic visible and near-infrared cameras (e.g. MegaCam and WIRCam at CFHT, or OmegaCam on the VST and VIRCam on VISTA at ESO/Paranal). TERAPIX is located at IAP (Institut d'Astrophysique de Paris, website http://terapix.iap.fr).} T0007 release of finely calibrated stacks and catalogues and photometric redshift data are publicly available and can be downloaded from the Canadian Astronomy Data Centre (CADC).
We use the version of the photo-z catalogue consisting of 4613209 sources where the overlapping regions between the observing tiles have been removed through a S/N criterion, and therefore multiple objects have already been removed.
We remove from the sample all bright objects with bad photometric redshift measurements, in order to avoid  high levels of contamination from spurious sources, such as saturated stars.
Observed magnitudes have been corrected for zero-point offsets that have been computed using spectroscopic redshifts by comparing the observed and modelled fluxes. The values of the zero-point corrections depend on the band of observation and band are given in Table \ref{zeropoint_corr}. 

\begin{table}
\begin{center}
\begin{footnotesize}
\begin{tabular}{cccccc}
u & g & r & i & y & z\\
\hline
0.00978 & -0.04726 & -0.02308 & -0.00567 & -0.01864 & 0.06455\\
\end{tabular}
\caption{Zero-point corrections for  the CFHTLS-W1 field. Offsets need to be subtracted from each band.\label{zeropoint_corr}}
\end{footnotesize}
\end{center}
\end{table}

No photometric redshift estimates are available for the Field A and Field B regions.
For a common subsample of galaxies with the W1 Field, we verified that the magnitudes of fields A and B were already corrected for zero-point offsets.

A unique photometric catalogue is created eliminating the overlap regions between W1, Field A and Field B and the final number of sources is 4508438. 

\begin{figure}
\begin{center}
\includegraphics[scale=0.35,clip, trim=8 8 15 15]{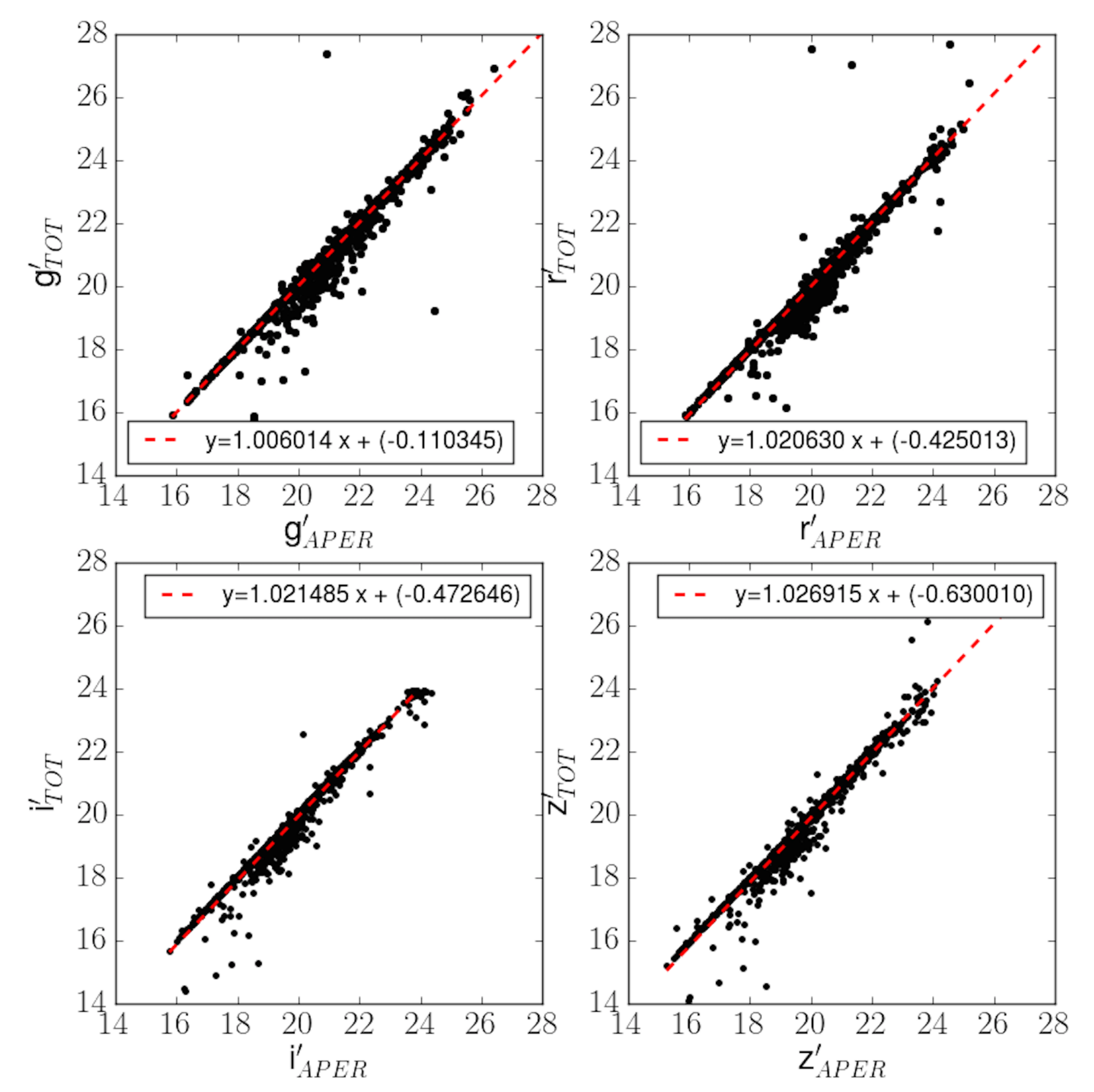}
\caption{Relation between aperture and total magnitude for the SF photometric catalogue (see text for details). Each panel refers to a different band; the red line is the linear fit used to convert aperture magnitudes into total values for the whole SF catalogue.
\label{polyfit_MagAper_Tot}}
\end{center}
\end{figure}

Finally, we also add new photo-z measurements in the W1 Field (Sotiria Fotoupoulou, private communication, hereafter SF catalogue). 
This catalogue contains aperture magnitudes in $g^\prime$, $r^\prime$, $i^\prime$, $z^\prime$, $J^\prime$, $H^\prime$, $K^\prime$ bands for 4887 galaxies. Using a common subsample of the SF catalogue and of the CFHTLS W1 catalogue, we derive the linear fit between aperture magnitudes from SF and total magnitudes from CFHTLS for each of the filters $g^\prime$, $r^\prime$, $i^\prime$, $z^\prime$: the offsets and slopes of the relations are written in each panel of Figure \ref{polyfit_MagAper_Tot} and are used to convert aperture magnitudes into total values for the 4887 matched objects. We note that the number of galaxies belonging to this sample that will be included in the scientific analysis presented in this paper and in the released catalogue is negligible (0.8\%).

We compute errors on total magnitudes combining in quadrature the mean error on total magnitudes calculated in 0.5 magnitude bins and the root mean square (rms) of the aperture-to-total magnitude relation shown in Fig. \ref{polyfit_MagAper_Tot}, calculated using the same binning in magnitude. 
The total number of sources with  photometric  information is 4513325.

We note that all magnitudes used are Sextractor \verb!MAG_AUTO! magnitudes \citep{BertineArnouts1996}  in the AB system corrected for Milky Way extinction according to \cite{Schlegel1998}.
Finally, we note that photometric redshift estimates are not used in the following, because all galaxies in the sample considered in the scientific analysis have spectroscopic redshift measurements.

\subsection{Spectroscopic database} \label{spec_data}

The galaxy spectroscopic information 
is hosted in the CeSAM (Centre de donn\'eeS Astrophysiques de Marseille) database in Marseille.\footnote{http://www.lam.fr/cesam/} The database contains data for both the XXL G\&C and the galaxies in the same area.
In addition to some XXL dedicated observing runs (XXL Paper XX), many other surveys have observed the galaxies in this field and the database includes them all. In particular, all redshifts from the VIMOS Public Extragalactic redshift survey, covering the redshift range $0.4<z<1.2$, were made available for this analysis prior to the recent public release \citep{Scodeggio2016}. As a result, there is a wide variety of spectra of different quality and origin to deal with. 
The final spectroscopic data release (CeSAM-DR2) is public and can be downloaded directly from the database; the complete list of the surveys and observing programmes included is given in Table \ref{list_CeSAM_dr1}. The samples included in the table refer to both XXL fields; however, as we focus here only on the XXL-N, we present the results only for this region.
\begin{table*}
\begin{center}
\begin{tabular}{l|l|l|l}

SpecOrigin in the  &\multirow{2}{*}{ Parent survey and References} &\multirow{2}{*}{Field} & \multirow{2}{*}{Type}\\
spectroscopic database &  & &\\
\hline
AAT\_AAOmega & \cite{2016PASA...33....1L} - XXL Paper XIV& S & PI XXL\\
AAT\_AAOmega\_GAMA & GAMA, Baldry et al. (2017, submitted) & N & E\\
AAOmega2012 & XXL Paper XX & N & PI XXL\\
Akiyama & \cite{2015PASJ...67...82A} & N & E\\
Alpha\_compilation & \cite{2011AA...526A..18A}& N & PI XMMLSS + E\\
ESO\_Large\_Programme & XXL Paper XX & N+S & PI XXL\\
LDSS03 & \cite{2011AA...526A..18A}& N & PI XMMLSS\\
Magellan & XXL Paper XX & N & E (XXL agreement)\\
Milano & \cite{2011AA...526A..18A}& N & PI XMMLSS + E\\
NED & & N+S & E\\
NTT & \cite{2011AA...526A..18A}& N+S & PI XMMLSS\\
SDSS\_DR10 & SDSS, \cite{2014ApJS..211...17A}& N & E\\
Simpson & \cite{2012MNRAS.421.3060S}& N & E\\
SNLS & Balland et al. (2017, in preparation)& N & E\\
Stalin & \cite{2010MNRAS.401..294S}& N & E\\
Subaru & \cite{2015PASJ...67...82A}& N & E\\
VIPERS/XXL and VIPERS\_2DR & VIPERS, \citep{Scodeggio2016} & N & E\\
VVDS\_Deep & VVDS, \cite{2005AA...439..845L}& N & E\\
VVDS\_UD & VUDS, \cite{2015AA...576A..79L}& N & E\\
XMMLSS & \cite{2011AA...526A..18A}& N & PI XMMLSS\\
WHT & \citet{Koulouridis2016} - XXL Paper XII & N & PI XXL\\
\end{tabular}
\caption{Surveys included in the first release of the CeSAM XXL database and contributing to our galaxy sample.  Entries in the first column are reported as they appear in the SpecOrigin column in the original database, and in the second column they have been grouped into main surveys and observing programmes relative to a given instrument or telescope. The \emph{Field} column indicates which XXL area is covered by the survey (North (N), South (S), or both), and the \emph{Type} column indicates the source of the data: E (External), PI (XXL or XMMLSS PI). \label{list_CeSAM_dr1}}
\end{center}
\end{table*}

The sample contains 134604 sources;
25421 of them refer to multiple observations of the same objects from different surveys, but the multiple measurements are not flagged in CeSAM.
To remove the duplicates, two different selection criteria are defined, both based on sets of priorities on observational properties of galaxies.
The first set of priorities regards the origin of the considered spectrum (the {\it SpecOrigin} column in the database catalogue). The different surveys are divided into three classes of priority ({\it origin flag}: 1, 2, 3): the smaller the value the higher the priority.
The list of the surveys with their attributed {\it origin flag} is given below:
\begin{enumerate}
\item (AAT\_AAOmega, entirely in the South), AAT\_AAOmega\_GAMA, ESO Large Programme, FORS2\_AAOmega, NTT, WHT, XMMLSS, SDSS\_DR10.
\item VIPERS/XXL, VVDS\_UD, VVDS\_deep.
\item Akiyama, Alpha\_compilation, LDSS03, Milano, NED, SNLS, Simpson, Stalin, Subaru, Magellan.
\end{enumerate}

The second  set of priorities is given on the basis of the reliability of the redshift measurement, as given by each survey (the $z_{flag}$ column in the database catalogue). All the possible values assumed by this flag in the different surveys are grouped into five classes ({\it quality flag}: 0, 1, 2, 3, 4): the higher the value the higher the precision and reliability of the redshift estimate. 
We list in the following all the original {\it flags} as they are in the CeSAM XXL spectroscopic database and the corresponding {\it quality flag} (the first number in the list) as they are in the final catalogue that is released in this paper (Appendix \ref{catalog}).
\begin{enumerate}[start=0]
\item  -99.99 ($z_{flag}$< -13.0 in the routine), 0, 20;
\item  1, 11, 21, 31, 311, -11;
\item  2, 9, 12, 19, 22, 29, 32, 39, 312, 319, 219, 75;
\item  3, 13, 23, 33, 313;
\item  4, 5, 14, 24, 34, 314.
\end{enumerate}

The selection for multiple measurements is then based on a consequential criterion that considers both  priorities: the redshift of the entry with smaller {\it origin flag} is adopted and, if more entries have the same  {\it origin flag}, the {\it quality flag} is considered, giving priority to the largest value.

Out of 25421 objects, 10165 with multiple redshift are selected using this method. In a further 3123 cases both flags coincide: for these, one spectrum is selected interactively and 1158 single objects are finally included in the catalogue. 

The `cleaned' spectroscopic catalogue is the ensemble of the catalogue of single spectra found in the parent catalogue (109183 sources, with {\it origin flag} =1 independently of their actual {\it SpecOrigin} and {\it quality flag} =400), of the zflag/SpecOrigin selected objects (10165 sources), and of the hand selected objects (1158 sources), and therefore it contains 120506 galaxies.
Overall, the uncertainties on the galaxy redshift in the database vary from 0.00025 to 0.0005, computed from multiple observations of the same object and depending on the sample used (more details on the XXL spectroscopic database are given in XXL Paper XX); we  consider the highest value in this range as the typical redshift error for all objects.

\begin{figure}
\begin{center}
\includegraphics[scale=0.35, clip=true,trim=8 0 40 20]{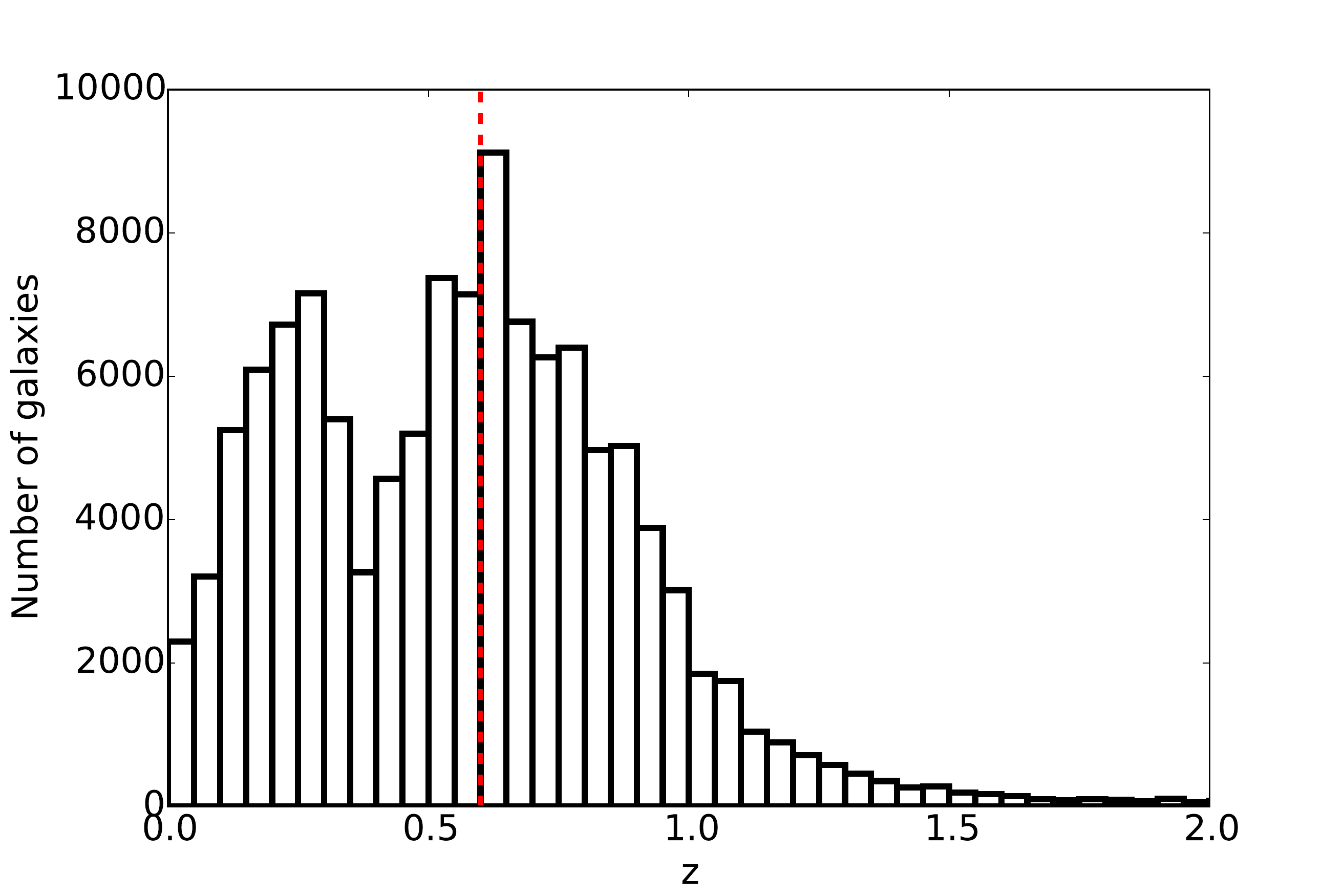}
\caption{Redshift distribution of the `cleaned' spectroscopic sample of galaxies (120506) from the CeSAM database in the XXL-N field. The vertical red dashed line corresponds to $z=0.6$, the maximum redshift of G\&C considered in this work.}
\label{z_distrib_CeSAM}
\end{center}
\end{figure}

The redshift distribution of the `cleaned' catalogue of galaxies
is shown in Fig. \ref{z_distrib_CeSAM}.

\begin{figure*}
\begin{center}
\includegraphics[scale=0.465]{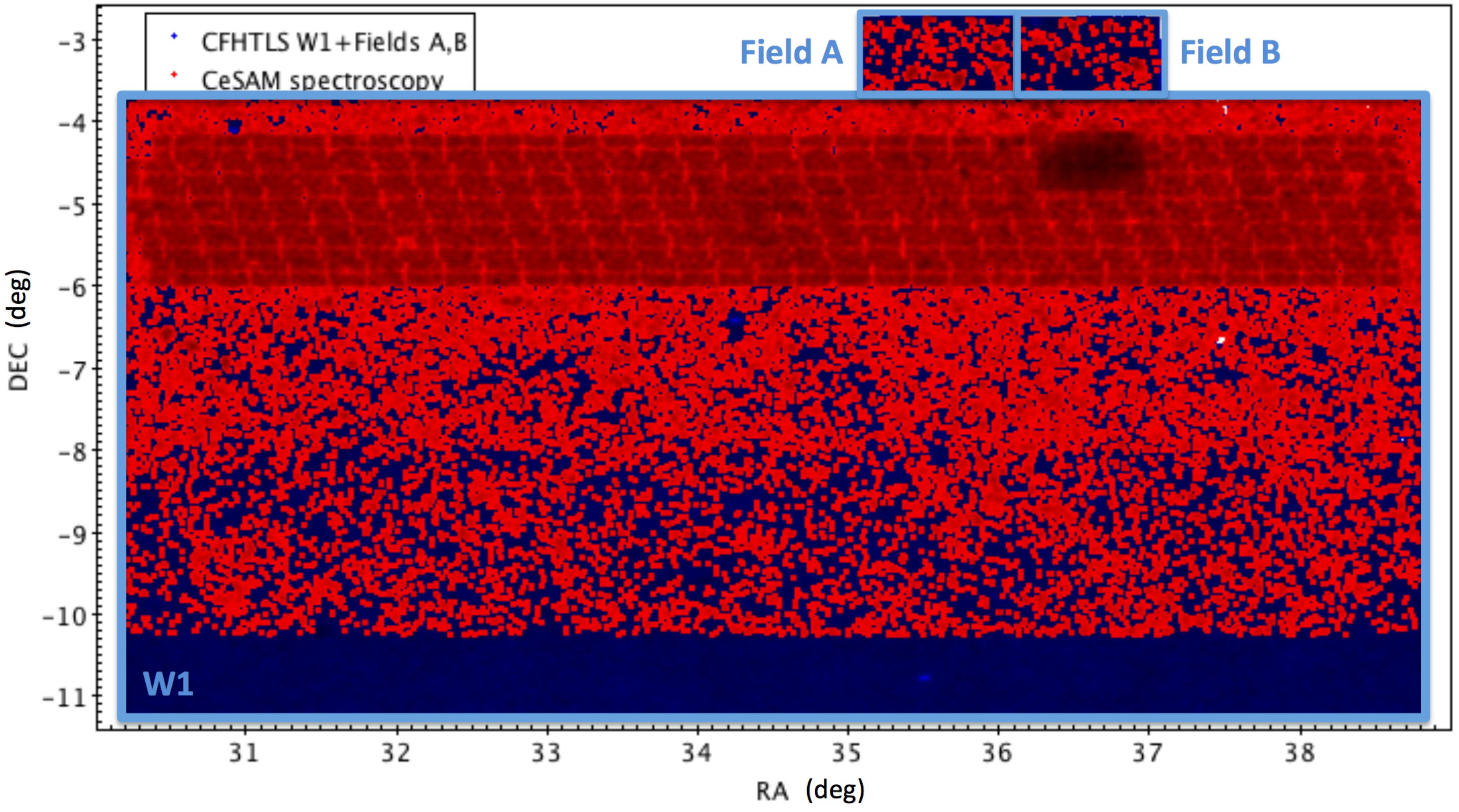}
\caption{CFHTLS W1, Fields A and B photometric catalogue (in blue) and CeSAM spectroscopic catalogue restricted to the photometric area (in red). Different signatures in the spatial distribution of the galaxies within the spectroscopic catalogue correspond to different sampling of the area performed by different surveys. In particular, the denser chess-board pattern in the upper part of the field ($-6 \lesssim DEC \lesssim -4.25 $) corresponds to the VIPERS data and the majority of the other red points are from the GAMA survey.}
\label{Spectra_photoz_catalogs}
\end{center}
\end{figure*}

%The main surveys contributing to the final spectroscopic catalog are:
%VIPERS with 51\% of all redshifts; GAMA with 27\% of all redshifts; SDSS  with  10\% of all redshifts; VVDS+VVUD  with 7\% of all redshifts; ESO + WHT observations  with $\sim 1\%$ of all redshifts.
As the last step, we combine the spectroscopic and the photometric catalogues.
We perform a match in coordinates between the two catalogues within 1 arcsec, obtaining 113732 galaxies. We exclude 
targets with redshifts $z\sim 0$ to avoid being contaminated by stars, and are left with 113223 galaxies.

In Figure \ref{Spectra_photoz_catalogs} the CFHTLS W1, Fields A and B photometric catalogue (4508438 sources) is shown together with the spectroscopic sample in the same region in the sky (114450 sources). We note that no redshift cut was applied in the spectroscopic catalogue shown in this figure.

\subsection{Spectroscopic completeness}
\label{sec_spec_compl}

The spectroscopic completeness of a sample is commonly defined as the ratio between the number of reliable spectroscopic redshifts in a given region and the total number of galaxies within it (i.e. the number of galaxies in the photometric catalogue). 
In principle, the completeness ratio depends on the sampling of the spectroscopic surveys in different regions of the sky (being our spectroscopic catalogue an heterogeneous ensemble of data coming from different surveys), the observed magnitude, and the colour of galaxies.
In order to deal with the first two factors mentioned above, we divide the XXL-N field into three stripes (arbitrarily named) according to the spatial distribution of the surveys:
\begin{itemize}
\item C-A: Number of galaxies in the spectrophotometric database = 3784, number of galaxies in the photometric database = 5292. Completeness =71.5$\pm0.8$\% (Poissonian error);
\item C-B: Number of galaxies in the spectrophotometric database = 15494, number of galaxies in the photometric database = 19944. Completeness =77.7$\pm0.4$\%;
\item C-C: Number of galaxies in the spectrophotometric database = 2497, number of galaxies in the photometric database = 8751. Completeness =28.5$\pm0.6$\%.
%\item C-A area: the north-east region corresponding to the beginning of the G02 GAMA region: RA=[30.17:38.83], DEC=[-4.1715:-3.72].
%\item C-B area: GAMA and VIPERS superposition area: RA=[30.17:38.83], DEC=[-6.0:-4.1715].
%\item C-C area: south-east area: RA=[30.17:34.0], DEC=[-8.0:-6.0].
\end{itemize}

\begin{figure*}
\begin{center}
\includegraphics[scale=0.45]{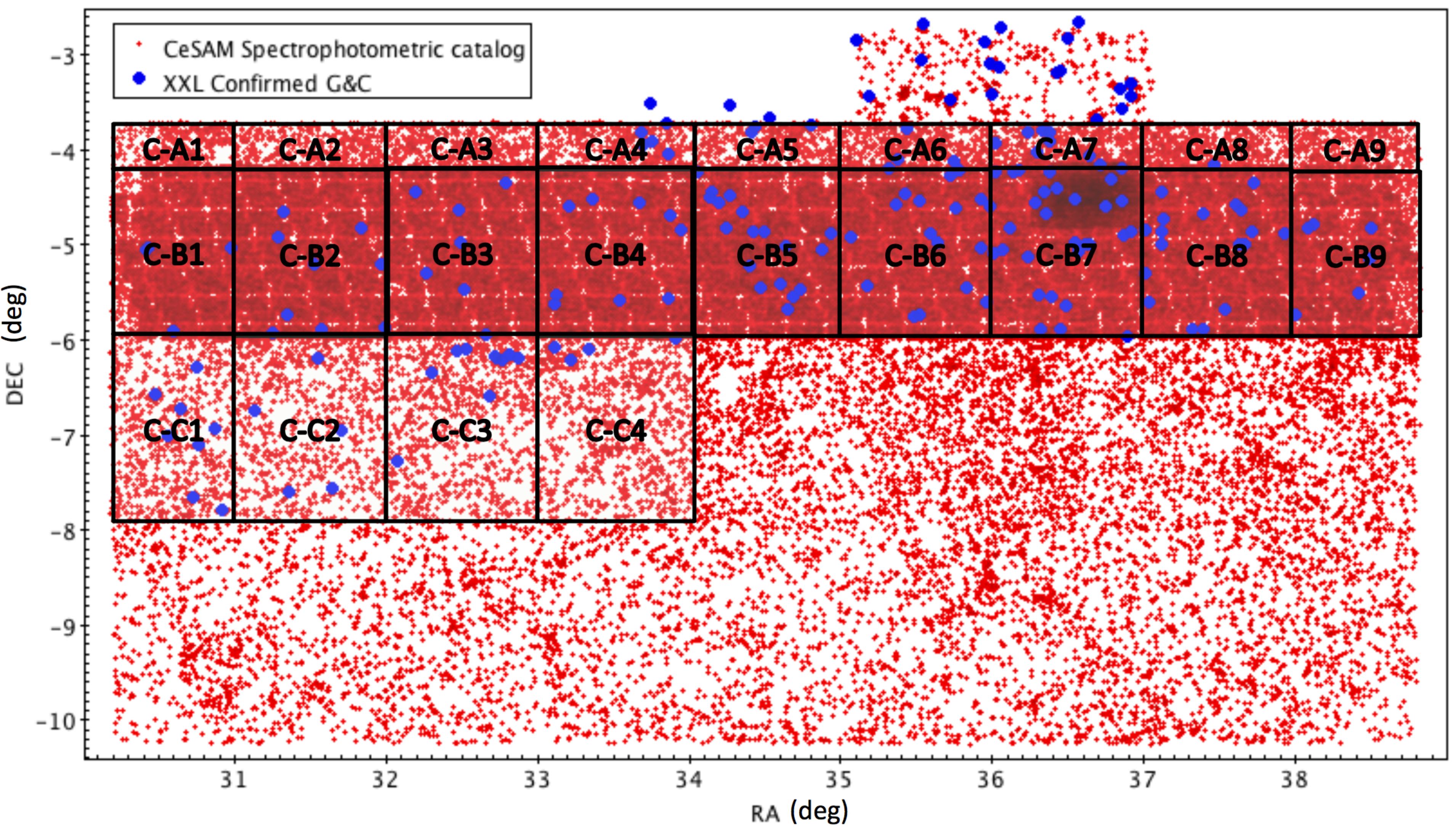}
\caption{XXL-N area. Red dots show the galaxies in the spectrophotometric sample used to compute the spectroscopic completeness (Section \ref{sec_spec_compl}) and blue dots represent X-ray confirmed G\&C. The regions in which the spectroscopic completeness has been computed are overplotted with small boxes.}
\label{compl_grid}
\end{center}
\end{figure*}

We then further subdivide the samples according to the position in the sky and the magnitude, creating a grid of 1.0 deg width both in RA and in DEC (for a total of 22 cells, see  Fig. \ref{compl_grid}), and considering intervals of 0.5 $r$-band observed magnitude.
We compute histograms of galaxies in each cell: the ratio of the spectroscopic to the photometric histograms gives the completeness in each region of the sky and in each magnitude bin within it. Completeness curves are obtained from the completeness ratio as a function of magnitude in each of the 22 cells.
Representative completeness curves are shown in Appendix \ref{app_compl}.

Considering the magnitude limited sample including 28096 galaxies with $r\leq 20.0$ (where the completeness drops dramatically, and which corresponds to GAMA observed magnitude limit $r=19.8$, see Appendix \ref{app_compl}), the completeness values for the three regions are as follows: 
\begin{itemize}
\item C-A: Number of galaxies in the spectrophotometric database = 4160, number of galaxies in the photometric database = 7491. Completeness =55.6$\pm0.8$\% (Poissonian error);
\item C-B: Number of galaxies in the spectrophotometric database = 17121, number of galaxies in the photometric database = 27923. Completeness =61.3$\pm0.4$\%;
\item C-C: Number of galaxies in the spectrophotometric database = 6815, number of galaxies in the photometric database = 13741. Completeness =49.6$\pm0.6$\%.
\end{itemize}

%Overall, within the sample, GAMA contributes with 92.8\% of all spectra, SDSS\_DR10 with 5\% of the spectra, and only a few spectra come from ESO LP and NTT XXL (39/28096), WHT (115/28096), VIPERS (129/28096) and VVDS (17/28096). 

\section{Definition of galaxy environment}
\label{membership}

\begin{figure}
%\begin{center}
\includegraphics[scale=0.33,clip, trim=8 30 25 70]{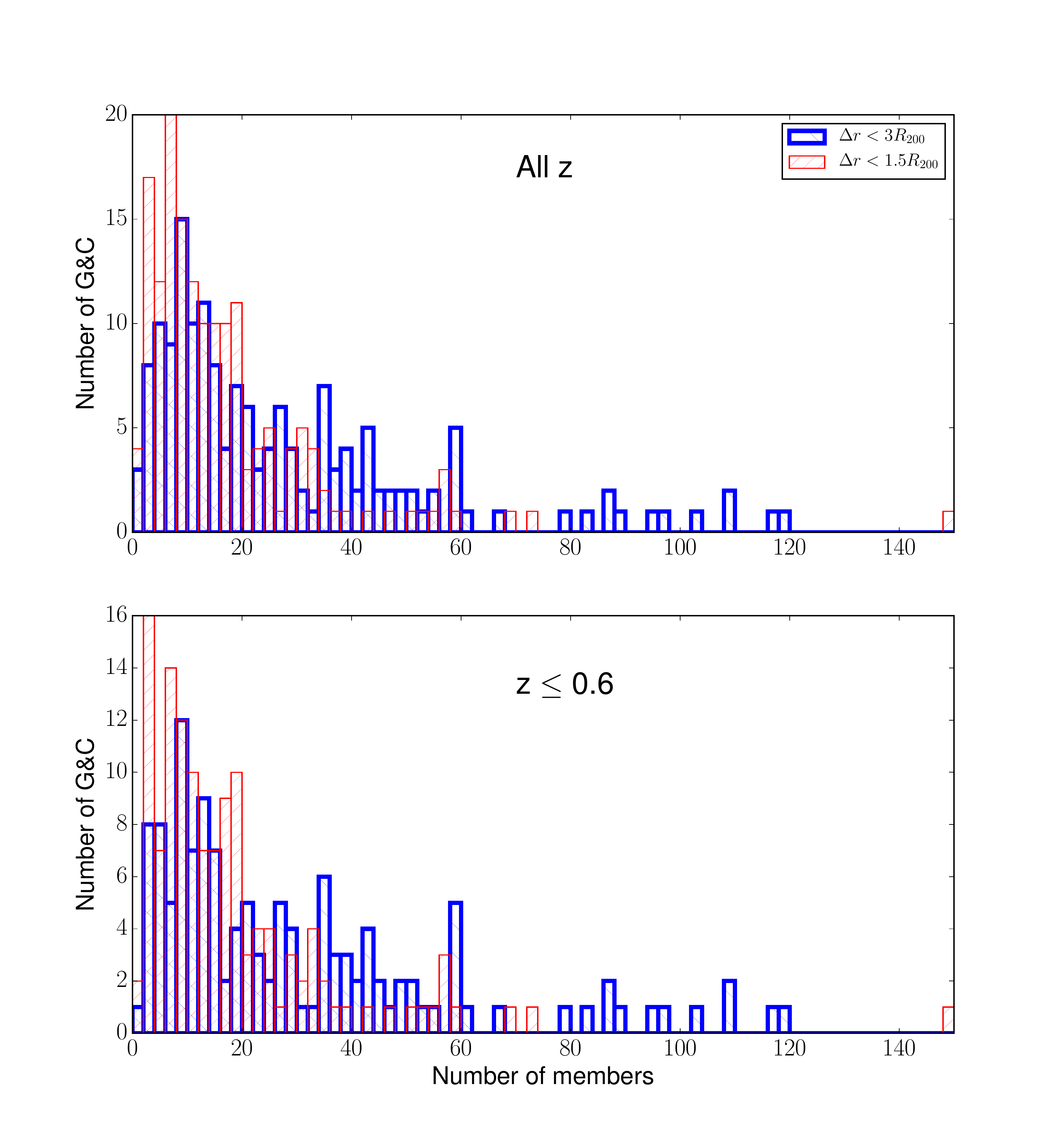}
\caption{Number of members in XXL-N G\&C at all redshifts (top panel) and in the 132 XXL-N G\&C at  $z\leq 0.6$ (bottom panel), assigned to structures as described in Sect. \ref{membership}. The 4180 members within $3 r_{200}$ (3619 at  $z\leq 0.6$) are plotted in blue; the 2656 members within $1.5 r_{200}$ (2284 at  $z\leq 0.6$) are plotted in red.}
\label{histo_members}
%\end{center}
\end{figure}

In order to determine which galaxies are part of our G\&C, we first need to compute the velocity dispersions of the structures.
We derive $\rm M_{200}$ from $\rm M_{500}$, using  the relations given by \citet{2006MNRAS.366..624B}, which is based on the concentration-mass relation from \citet{Dolag2004}:\footnote{In \citet{2016A&A...592A...4L}, hereafter XXL Paper IV, the relation from \citet{Duffy2008} is adopted instead.}
\begin{equation*}
 \frac{M_{200}}{M_{500}} =
  \begin{cases}
    1.30       & \quad \text{if } \quad 8 \times 10^{12}M_{\odot}<M_{500} \leq 5 \times 10^{13}M_{\odot} \\
    1.35  & \quad \text{if } \quad 5 \times 10^{13}M_{\odot}<M_{500} \leq 2 \times 10^{14}M_{\odot} \\
    1.40 & \quad \text{if } \qquad \qquad \qquad \quad M_{500} > 2 \times 10^{14} M_{\odot}
  \end{cases}
\end{equation*}

Then, using the virial theorem, we obtain  the velocity dispersion $\rm \sigma_{200}$ from $\rm M_{200}$ using  the relation given in \cite{2006ApJ...642..188P} (originally given in \citealt{2005ApJ...630..206F}):
\begin{equation}
\sigma_{200} = 1000 \, {\rm km \, s^{-1}} \cdot \left(\frac{M_{200}}{1.2 \cdot 10^{15} M_\odot} \cdot \sqrt{\Omega_{\Lambda} + \Omega_0(1+z)^3} \cdot h\right)^{1/3}
\label{sigma_200}
\end{equation}

We stress that these velocity dispersions are derived from X-ray-based mass estimates and are more reliable than values obtained from galaxy redshifts, especially for structures containing just few members. 

We derive r$_{200}$ from r$_{500}$, simply dividing the latter by 0.7, according to the relation adopted in \citet{Ettori2008}.

A galaxy is considered member of a G\&C if its velocity  $v_{gal}  =  c  (z_{gal} - z_{G\&C})/(1+z_{G\&C})$ lies within $\pm 3 \sigma_{200}$ where $z_{gal}$ is the spectroscopic redshift of the galaxy and  $z_{G\&C}$ the redshift of the G\&C, and if its projected distance from the G\&C centre is $< 3 \, r_{200}$ (`outer membership' region), or its projected distance from the G\&C centre is  $ < 1.5 \, r_{200}$ (`inner membership' region). 

This method identifies 4180 (2656) members within 3 (1.5) $r_{200}$ in the cleaned spectroscopic sample. We note that some galaxies are assigned multiple memberships, that is they can belong to different G\&C. This happens when two or more G\&C are physically close in space.  Specifically, when we consider the outer membership, 13\%/2\%/0.9\%/0.6\% of the galaxies in our sample simultaneously belong to two/three/four/five G\&C. No galaxy belongs to more than five G\&C.

Figure \ref{histo_members} shows histograms of the populations of G\&C as functions of the number of members. Separate panels show the full sample and those G\&C at $z\leq 0.6$, and the effects of using the inner and outer membership criteria are illustrated. We will use the outer membership criterion in all the analyses in the present paper. We find 95\% of all G\&C to have at least three spectroscopic members, and 70\% have at least ten members.

The field sample is defined as the ensemble of all galaxies not belonging to any G\&C.
We note that, due to the detection limit and sensitivity of X-ray observations, G\&C selected are preferentially dense structures, particularly at higher redshift where only higher mass G\&C are detected. As a consequence, galaxies that belong to lower mass structures at these redshifts (i.e. groups which are below the adopted X-ray thresholds, including c3 objects) are included in the field sample, and can in principle contaminate it.

We verified that the completeness strategy described in the previous section does not depend on galaxy colour or on environment (field versus structure), validating our adopted procedure. Appendix \ref{app_compl} shows the details of this additional analysis.

\section{Stellar masses}
\label{masses}

We compute stellar masses for all galaxies in the spectrophotometric sample using LePhare. This code was developed mainly to compute photometric redshifts (see Sect.\ref{photo}), but the code can also compute physical properties of galaxies such as stellar masses and star formation rates (SFR), and the spectroscopic redshift can be used as an input fixed parameter in order to improve the quality of the physical outputs.

Taking as inputs at least two observed magnitudes
and spectroscopic redshifts, the program proceeds through different phases:
\begin{itemize}
\item Creation of libraries (Stellar, QSO, Galaxy): we use the default LePhare lists for the stellar and QSO libraries, and the galaxy library was built from \cite{2003MNRAS.344.1000B} models, which consider an exponentially declining star formation histories SFH $\propto$ $ 1/\tau \cdot \exp(-t/\tau)$, where $t$ is the time since the age of formation, set at 13.5 Gyr, and $\tau$ is the decay time, i.e. the timescale of the star formation process;
\item Creation of theoretical absolute magnitudes from the input libraries: this phase requires several parameters that have to be set in order not to exceed the dimension of the library, defined as:\\
\begin{center}
$\rm Number \, of \, models \, \times \, Number \, of \, ages \, \times \, Number \, of \, SFHs \, \times \, Number \, of \, z-steps \, \times \, Number \, of \, extinction \, laws \, \times \, Number \, of \, E(B-V)$.\\
\end{center}
Out of the whole library of available models, 27 \cite{2003MNRAS.344.1000B} models were selected, that is nine SFHs with different $\tau$ values for each of the three metallicity values: Z=0.004, Z=008, Z=$Z_{\odot}$ = 0.02. We consider all the possible values for $\tau$ with solar metallicity, we choose a redshift step of 0.02 up to redshift 1.8 and the following values of E(B-V): 0.0, 0.05, 0.1, 0.15, 0.2, 0.25, 0.3, 0.35.
We extinguish galaxy models using the \cite{Calzetti2000} extinction law for $\rm \tau > 2$ star formation histories (i.e. extinguishing all star forming galaxies which show active star formation up to $\rm z \sim 0.5$, $\sim$8 Gyr from the Big Bang). According to the exponentially declining SFH model adopted, this corresponds to the SFH of a galaxy whose star formation activity is negligible at $\rm z \leq 0.5$. Galaxies with more rapidly declining SFHs are not extinguished by the code.

\item Application of the Photo-z code that performs the $\chi^2$ fitting analysis between the template and observed flux. The code can be used to extract physical information on individual galaxies such as absolute luminosities, stellar masses, SFR.
\end{itemize}

We take the stellar mass value as being the output stellar mass from the maximum likelihood (ML) analysis (MASS\_MED), which has an associated error, instead of the stellar mass corresponding to the minimum $\chi^2$, which is computed for all galaxies having a measured magnitude at least in one band. The mean and median error on the stellar mass estimates are computed from the minimum and maximum stellar mass estimate of LePhare  (MASS\_INF and MASS\_SUP), and are respectively 0.3 dex and 0.2 dex.

The program successfully computed stellar masses for 108151/113223 galaxies  (95.5\% of the sample). In the other cases the code did not converge, because of an insufficient number of magnitude bands given as input or a bad redshift estimate for the galaxy (e.g. if {\it quality\_flag} is 0 or 1).

\begin{figure}
\begin{center}
\includegraphics[scale=0.30]{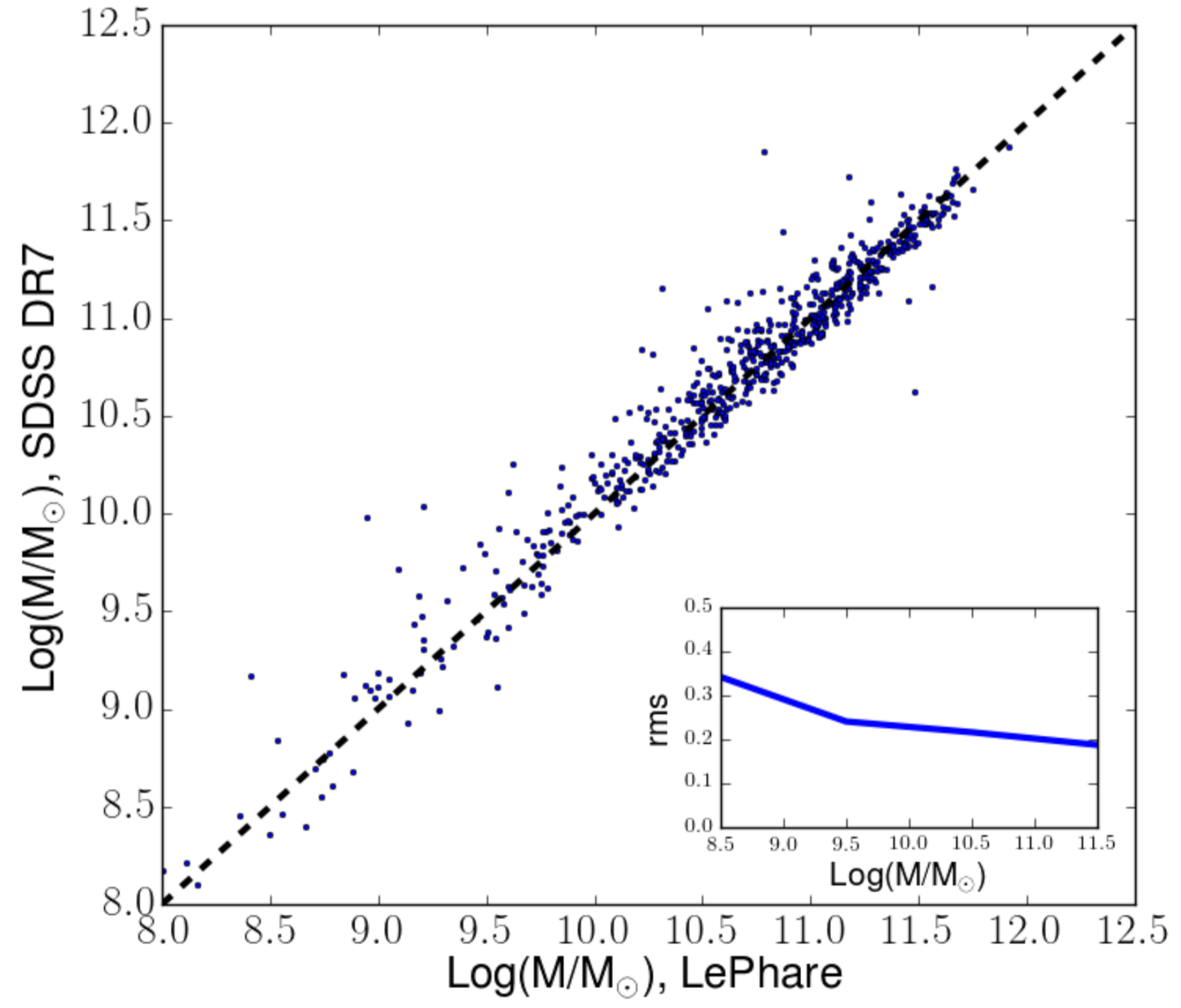}
\caption{Comparison between the stellar mass values computed with LePhare (this work) and the stellar masses from the SDSS DR7. The inset shows the root mean square (rms) as a function of mass between the two estimates.}
\label{comp_sdss_LePhare}
\end{center}
\end{figure}

To test our mass estimates, we compare our values with the masses obtained fitting the photometry following the methodology presented in \citet{2003MNRAS.341...33K,2007ApJS..173..267S,2004MNRAS.351.1151B}, for a subsample of 740 galaxies galaxies in common with the  SDSS DR7.
The comparison shown in Figure \ref{comp_sdss_LePhare} shows a good agreement between the stellar masses. The dispersion of the relation as a function of the stellar mass, given in the inset, is comparable with the error on stellar masses computed from LePhare. This test confirms the reliability of the LePhare configuration adopted.

We then compute again the spectroscopic completeness considering only the galaxies with stellar mass estimates. This completeness will be used in the following scientific analysis.

\subsection*{Stellar mass completeness limits}

\begin{figure}
\begin{center}
\includegraphics[scale=0.47]{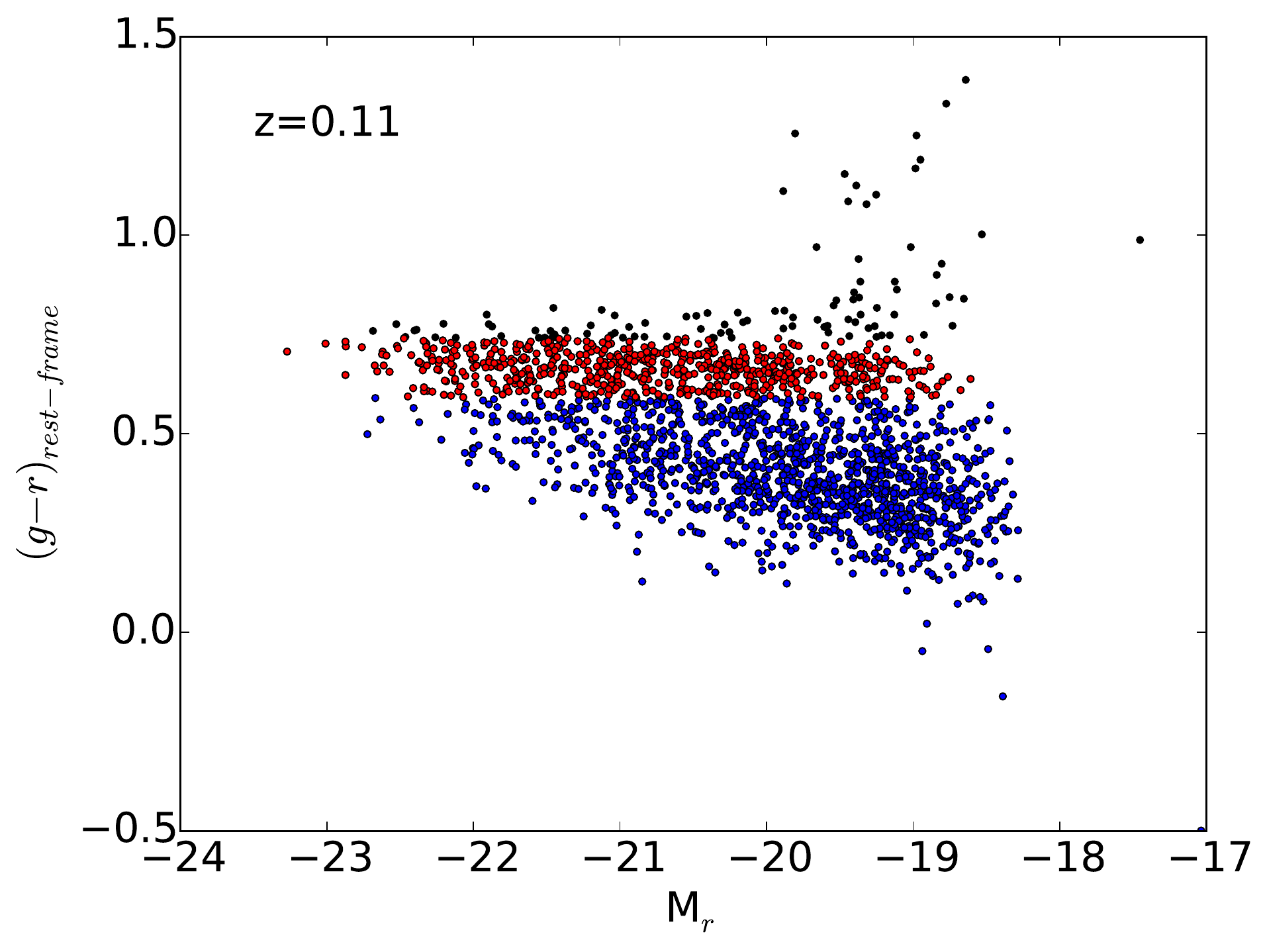}
\caption{Colour-magnitude diagram for galaxies in the redshift bin centred around $z=0.11$. Black points are the 5\% reddest galaxies excluded as outliers in the first step of the mass completeness limit computation. Red points represent the 0.15 width colour stripe used to define the absolute magnitude limit starting from the rest-frame colour limit (see text). Blue points are all other objects which are not considered for the mass limit computation. \label{cmd_example}}
\end{center}
\end{figure}

\begin{figure}
\begin{center}
\includegraphics[scale=0.47]{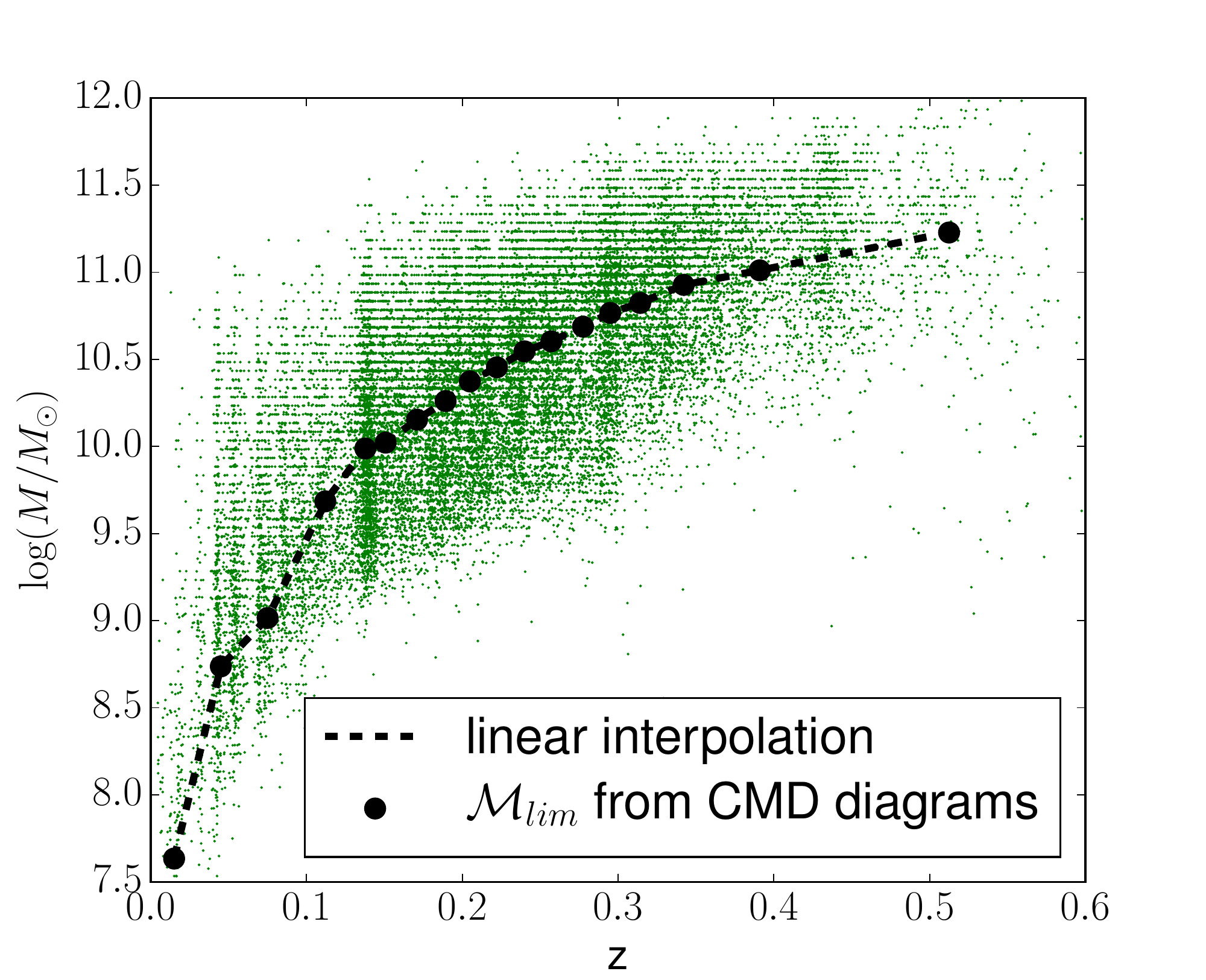}
\caption{Stellar mass completeness limit as a function of redshift. The black points represent the measured limit (see text). The black dotted line is the linear interpolation to the points. Green dots represent the entire galaxy sample.}
\label{Mass_z}
\end{center}
\end{figure}

The magnitude limit of $r=20$ can be translated  into a stellar mass limit. 
%This limit is defined as the mass of the galaxy with the reddest rest-frame color and with an absolute magnitude that corresponds to the observed magnitude limit at its spectroscopic redshift.
This limit is strongly redshift dependent, so to compute it we divide our entire redshift range into several intervals.
We consider fixed redshift bins of $\Delta z = 0.03$ up to $z=0.09$, and bins with a fixed number of galaxies (2000) in the redshift range $0.09 < z \leq 0.6$. When computing  stellar mass limits, we do not separate galaxies according to their environment.

Considering only the galaxies entering the magnitude limited (r=20.0) sample and focusing on one redshift interval at a time, we compute the mass limits as follows:
\begin{itemize}
\item We build the $(g-r)_{rest-frame}$ vs. $r$ rest-frame colour-magnitude diagram for galaxies entering the sample.
Excluding the 5\% reddest galaxies to eliminate outliers, we define as rest-frame colour limit $(g-r)_{rest-frame,lim}$ the colour of the reddest galaxy in the sample;
\item We identify the so-called red sequence, selecting galaxies with $(g-r)_{rest-frame,lim}-0.15< (g-r)_{rest-frame}<(g-r)_{rest-frame,lim}$. We then define the absolute magnitude limit M$_{r,lim}$ as the absolute r-band magnitude of the faintest galaxy in the interval;
\item We derive the mass limit following \cite{Zibetti2009},
\begin{footnotesize}
\begin{equation}
\mathcal{M}_{lim,M_{\odot}}=-0.840+1.654(g-r)_{rest-frame,lim} + 0.4(M_{r,\odot} - M_r)
\end{equation}
\end{footnotesize}
where the absolute magnitude of the Sun is $\rm M_{r,\odot} = 4.64$.
\end{itemize}

As an example of the procedure, Fig. \ref{cmd_example} shows the colour-magnitude diagram for galaxies at $z=0.11$.

Finally, we use an interpolation method to obtain the mass limit at each desired redshift (Fig. \ref{Mass_z}).

\section{Final catalogue}\label{sp_cat}

The final catalogue  used in our analysis and made publicly available to the community at CDS contains all the properties described in this paper for the subsample of galaxies with $0<z\leq 0.6$, $r \leq 20$ and a derived stellar mass estimate. 
The released  sample is composed of 24336 galaxies,  both in the field and in G\&C, and the contribution of the different surveys is the following: 95\% of redshifts come from GAMA (23178 galaxies out of 24336), 3\% are from SDSS\_DR10 (763 galaxies out of 24336) and the rest come from ESO Large Programme + WHT XXL dedicated observational campaigns (0.5\%, 115 galaxies out of 24336) and VIPERS (0.2\%, 48 galaxies out of 24336).
The catalogue contains  the astrometry from CFHTLS, the redshift, the name of the parent catalogue/survey, the \textit{origin flag} and \textit{quality flag}  that were introduced in Sect. \ref{spec_data}, all the membership related quantities, absolute magnitudes, stellar masses and completeness values.
A detailed description of all the entries provided is presented in Appendix \ref{catalog}. 

Table \ref{datasample_all} presents some useful numbers characterising the sample. Overall, 22111 (29683 once weighted for incompleteness) galaxies enter the field sample, 2225 (3446 once weighted for incompleteness) galaxies enter the G\&C sample, which includes 132 structures. 

\begin{table}
\begin{center}
\begin{tabular}{c|c|c|c}
z & N$_{\textrm{G\&C}}$ & N$_{\textrm{gals}}$ in G\&C & N$_{\textrm{gals}}$ in the field 	\\

\hline
0.0-0.1 & 11 & 294 (393) & 2228 (3015) \\
0.1-0.2&  24  & 991 (1147) & 6125 (7868) \\
0.2-0.3&  37  & 634 (743) & 8134 (10708)\\
0.3-0.4& 22  & 212 (320) & 4206 (5791)\\
0.4-0.6 &  38 & 94 (191) & 1418 (2301)\\
\hline
total & 132  & 2225 (2794) & 22111 (29683)\\
\end{tabular}
\caption{Statistics of the sample at $r\leq 20$. Numbers are given according to the redshift bins given in Col. 1. Column 2 gives the total number of G\&C in each redshift bin, Col. 3 gives the number of galaxies in the G\&C, while Col. 4 gives the number of galaxies in the field. The numbers in parentheses are weighted for spectroscopic completeness.}
\label{datasample_all}
\end{center}
\end{table}

\begin{table*}
\begin{center}
\begin{footnotesize}
\begin{tabular}{c|c|r|r|r|r|r|r|r}
z & $\mathcal{M}_{lim, M_\odot}$  & \multicolumn{3}{c|}{N$_{\textrm{G\&C}}$} & \multicolumn{3}{c|}{N$_{\textrm{gals}}$ in G\&C} & N$_{\textrm{gals}}$ in the field 			\\
&	& all & $L_{XXL}^{500}<10^{43}$ erg/s&  $L_{XXL}^{500}>10^{43}$  erg/s & all &$L_{XXL}^{500}<10^{43}$ erg/s &  $L_{XXL}^{500}>10^{43}$  erg/s &\\

\hline
0.1-0.2& 9.6 & 21 & 16 & 5 & 920 (1116) & 420 (530) & 500 (586) & 4402 (6098) \\
0.2-0.3& 10.4 & 34 & 17 & 17 & 502 (751) & 182 (272) & 320 (479) & 4654 (6729)\\
0.3-0.4& 10.8 & 24 & 11 & 13 & 187 (351) & 85 (135) & 102 (216) &  2468 (4009)\\
0.4-0.6 & 11.0 & 38 & - & - & 141 (531) & - & - & 2595 (13188)\\
\hline
total && 117 &44& 35 & 1746 (3132) & 687 (937)& 922 (1281)& 14119 (30024) \\
\end{tabular}
\caption{Final sample used in the  analysis. 
Numbers are given according to the four redshift bins given in Col. 1. Column 2 gives the stellar mass limit. Columns 3-5 indicate the total number of G\&C, and the number of G\&C in the two luminosity bins in which the GSMF has been studied, which contain galaxies with stellar masses above the mass limit. The remaining columns indicate the number of galaxies above the stellar mass limit in G\&C, divided again into luminosity classes, and the number of galaxies in the corresponding field sample; the quantities in parentheses refer to the number of galaxies weighted for spectroscopic completeness.}
\label{datasample}
\end{footnotesize}
\end{center}
\end{table*}

\section{Galaxy stellar mass function}
\label{results}

In the previous sections we have built catalogues of galaxies in G\&C and in the field with $r$-band magnitude $\leq 20$. In this section we present a first scientific exploitation of the sample and characterise the stellar mass distribution, investigating its dependence on environment and redshift. 

In both environments, we divide galaxies into four broad redshift bins:
$0.1\leq \Delta z \leq 0.2$, $0.2 < \Delta z \leq 0.3$, $0.3< \Delta z \leq 0.4$, $0.4< \Delta z \leq 0.6$. We exclude the lowest bin ($0<z< 0.1$) because, as shown in Table \ref{datasample_all}, our catalogue includes only 11 groups at these redshifts and we do not have a representative sample of the general population at this epoch. 
Using the linear interpolation given in  Fig.\ref{Mass_z}, we assign to each redshift bin the  stellar mass completeness limit corresponding to the lower end of each interval. 

\begin{figure}
\begin{center}
\includegraphics[scale=0.45]{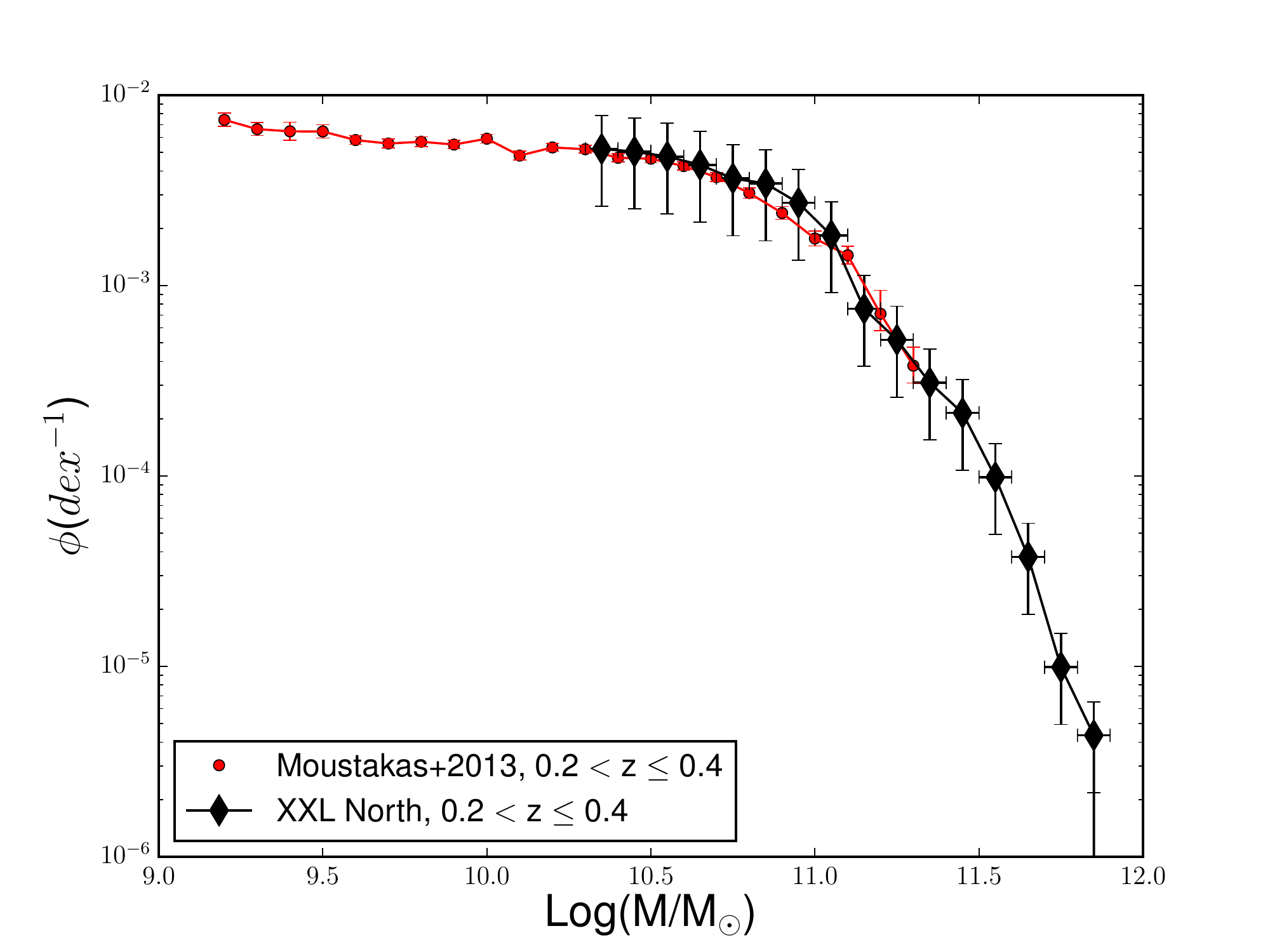}
\caption{Comparison between the galaxy stellar mass function of XXL-N field galaxies in the redshift range 0.2-0.4 and the stellar mass function derived in \cite{Moustakas2013} in the same redshift range. The original normalisation of \cite{Moustakas2013} was maintained and the values of the GSMF of XXL-N field derived in this work are scaled to theirs in the common mass range (see text). The survey is more sensitive to low-mass galaxies, but is smaller than XXL-N and does not probe the high-mass end of the galaxy population.}
\label{mf_XXL_Moustakas}
\end{center}
\end{figure}

\begin{figure*}
\centering
\includegraphics[scale=0.5]{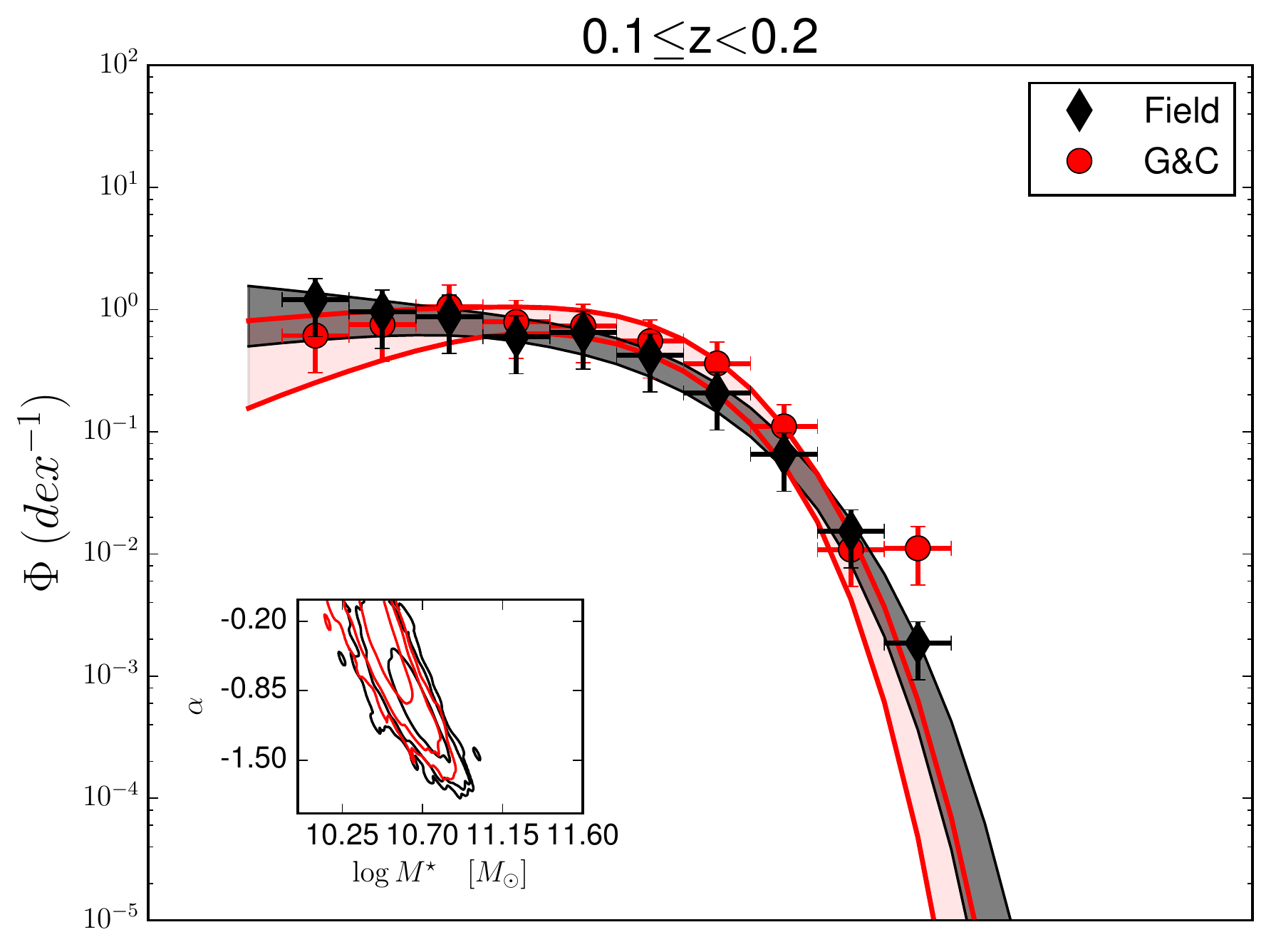}
\includegraphics[scale=0.5]{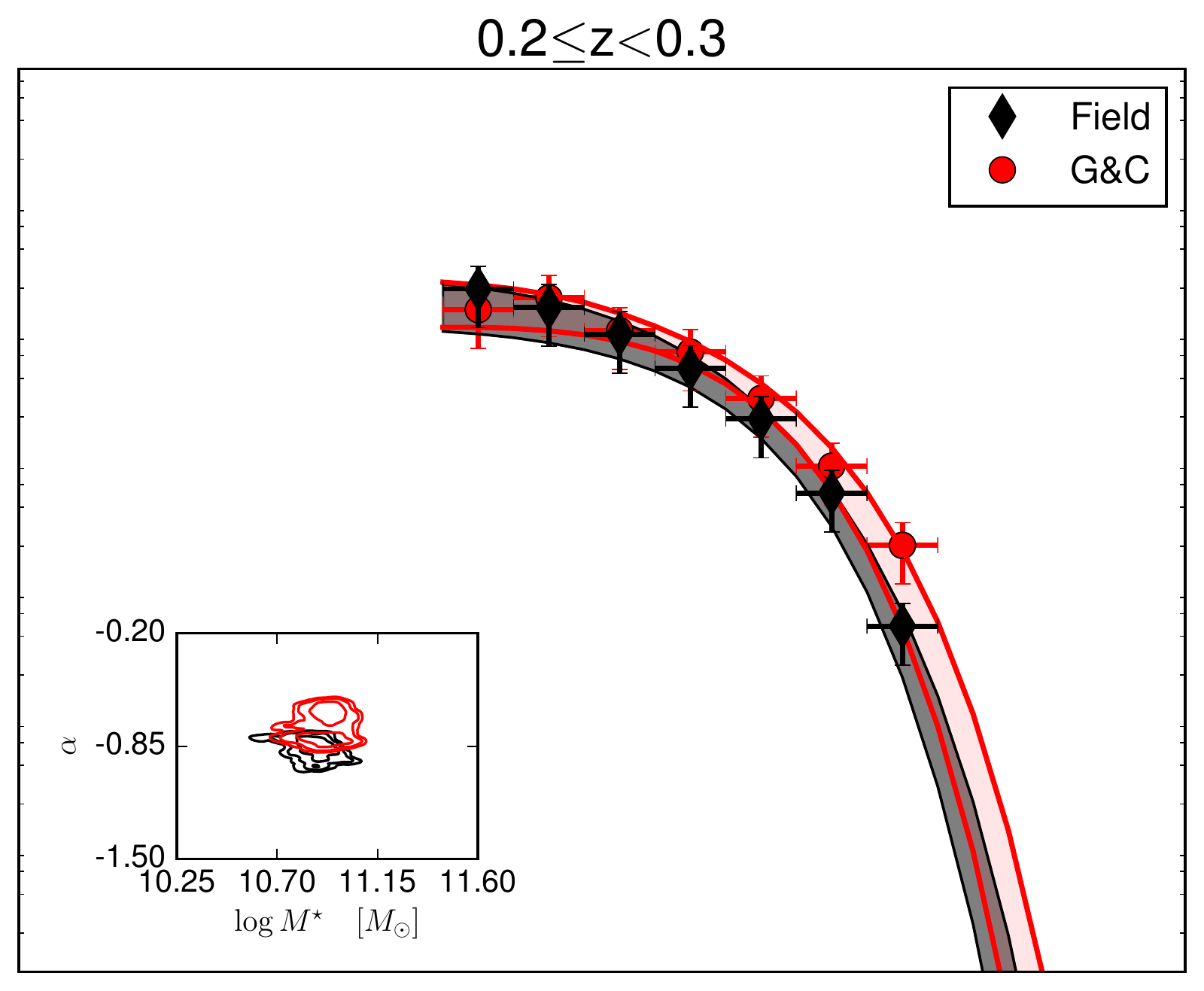}
\includegraphics[scale=0.5]{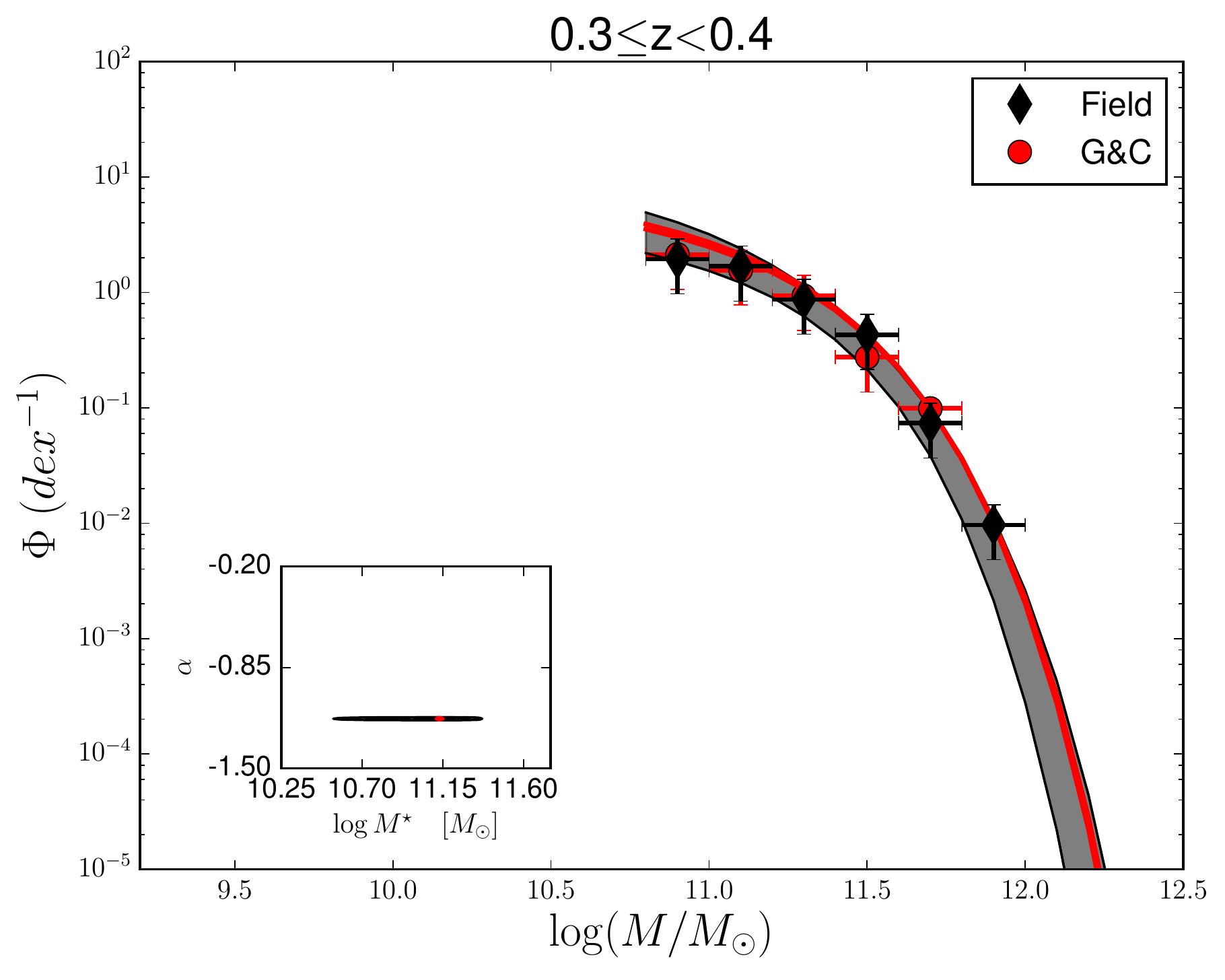}
\includegraphics[scale=0.5]{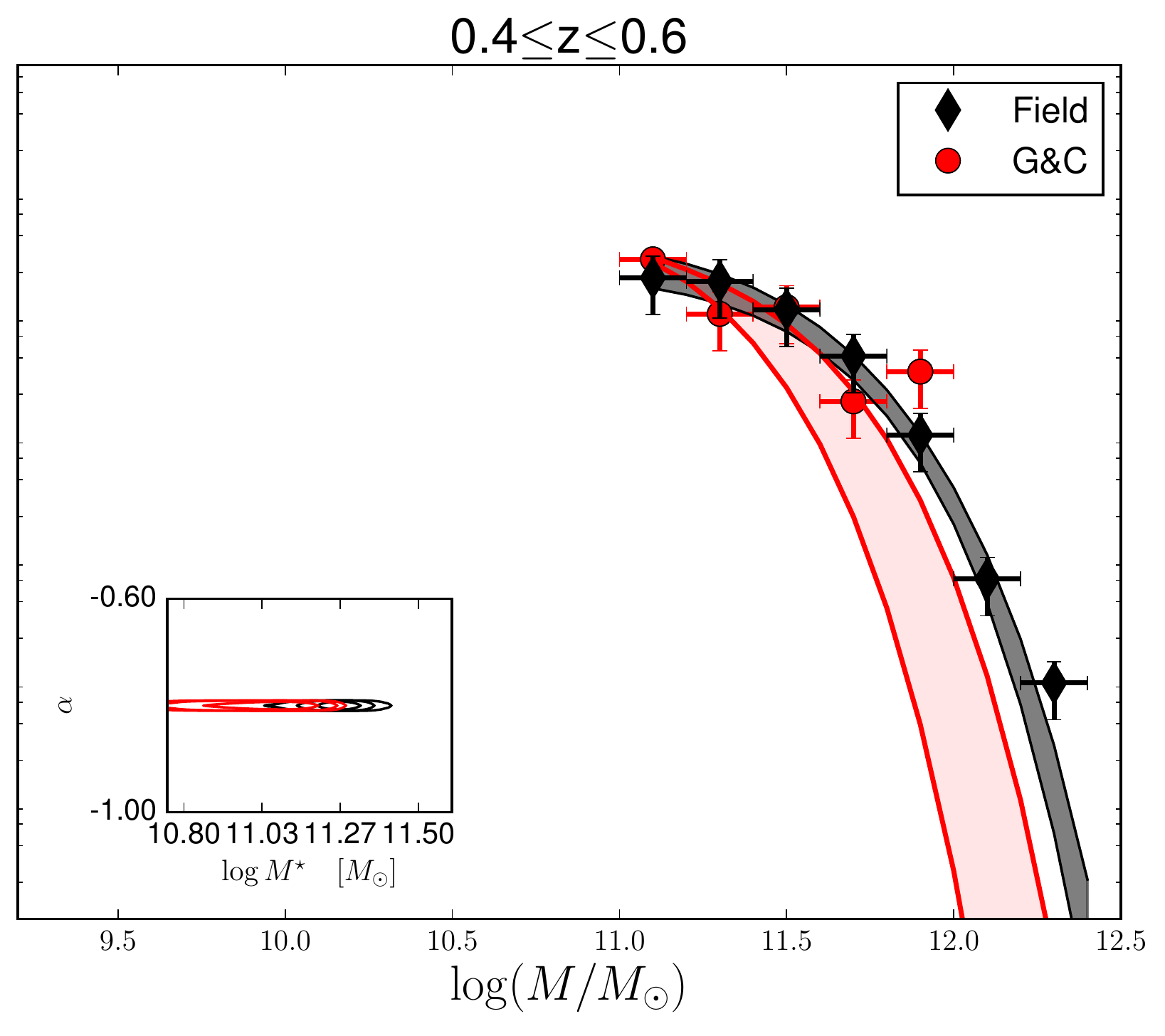}
\caption{Galaxy stellar mass function in different redshift ranges, as indicated in each panel, for galaxies in G\&C (red points) and in the field (black diamonds). Only points above the mass completeness limit are shown. Error bars on the x-axis show the width of the mass bins; those on the y-axis are derived from Poisson's statistics on the number counts together with the cosmic variance contribution. Schechter fit functions are also shown as shaded areas and follow the same colour scheme as the points. In the insets, 1,2,3 $\sigma$ contour plots on the Schechter fit parameters $\alpha$ and $M^\ast$ are also shown. At z$\geq$0.3, fixed values for the faint end slope $\alpha$ were set in order to perform Schechter fits.}
\label{mf}
\end{figure*}
   
 \begin{table}
\begin{center}
\begin{tabular}{l|r|r|r}
  \multicolumn{1}{c|}{z} &
  \multicolumn{1}{c|}{Environment} &
  \multicolumn{1}{c|}{$\alpha$} &
  \multicolumn{1}{c}{$\log(M^\star/M_\odot)$} \\
\hline
 \multirow{4}{*}{0.1-0.2} & Field & -0.8 $\pm$ 0.4 & 10.6 $\pm$ 0.1 \\
  & G\&C & -0.01 $\pm$ 0.50 &  10.4 $\pm$ 0.1 \\
  & G\&C High $L^{XXL}_{500}$ & 0.4 $\pm$ 0.7 & 10.3 $\pm$ 0.1  \\
  & G\&C Low $L^{XXL}_{500}$ & 0.1 $\pm$ 0.5 & 10.4 $\pm$ 0.1  \\
  \hline
   \multirow{4}{*}{0.2-0.3} & Field & -0.80 $\pm$ 0.05 & 10.87 $\pm$ 0.06 \\
  & G\&C & -0.59 $\pm$ 0.08 & 10.98 $\pm$ 0.03 \\
   & G\&C High $L^{XXL}_{500}$ & -0.87 $\pm$ 0.08 & 10.94 $\pm$ 0.04 \\
  & G\&C Low $L^{XXL}_{500}$ & -0.72 $\pm$ 0.06 & 11.07 $\pm$ 0.06\\
  \hline
   \multirow{4}{*}{0.3-0.4} & Field & -1.18 $\pm$ - & 11.1 $\pm$ 0.1\\
  & G\&C & -1.18 $\pm$ - & 11.125$\pm$ 0.007\\
   & G\&C High $L^{XXL}_{500}$ & -1.18 $\pm$ - & 11.1 $\pm$ 0.1\\
  & G\&C Low $L^{XXL}_{500}$ & -1.18 $\pm$ - & 10.69 $\pm$ 0.09 \\
  \hline
   \multirow{2}{*}{0.4-0.6} & Field & -0.8 $\pm$ - & 11.27 $\pm$ 0.05\\
  & G\&C & -0.8 $\pm$ - & 11.07 $\pm$ 0.14\\
\end{tabular}
\caption{Best-fit Schechter Function Parameters $\rm(M^\star,\alpha)$ for the GSMF in different environments and redshifts. For $z \geq 0.3$ we fixed $\alpha$ in our fits, therefore it does not have errors. At $z>0.4$, due to low number statistics, we cannot divide our sample into low and high $L^{XXL}_{500}$ G\&C.
\label{Schechter_param}}
\end{center}
\end{table}

\begin{figure*}
\centering
\includegraphics[scale=0.35]{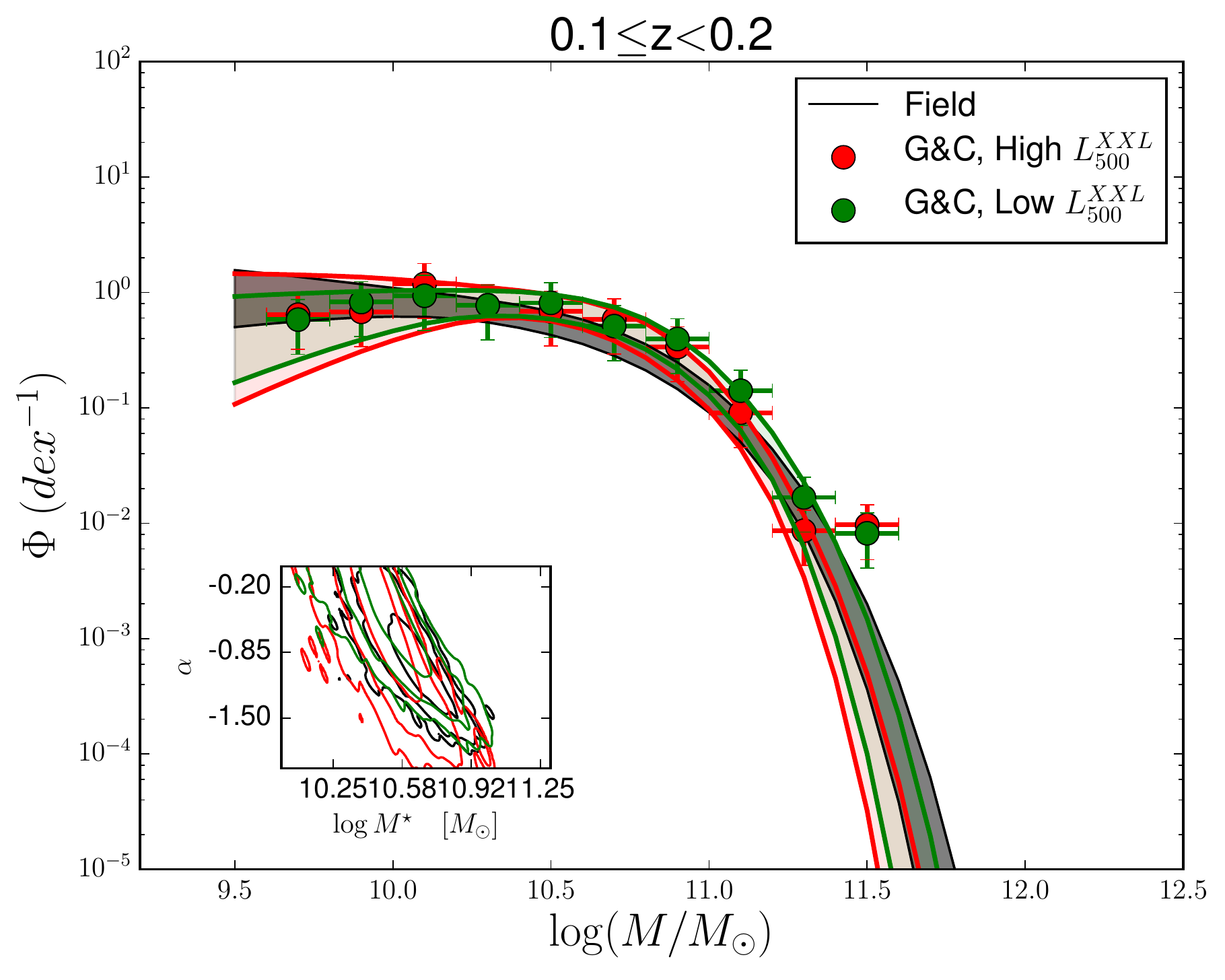}
\includegraphics[scale=0.35]{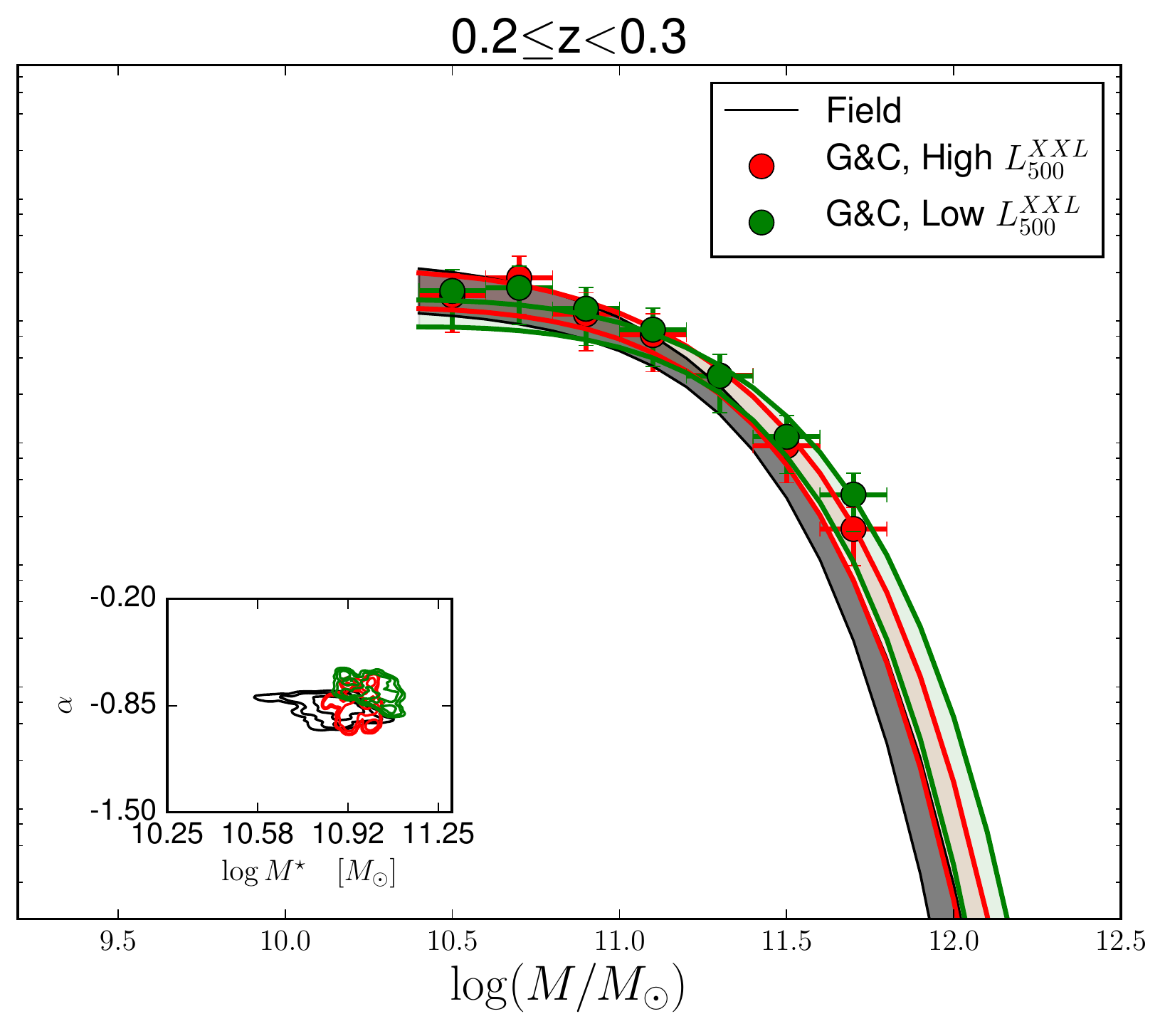}
\includegraphics[scale=0.35]{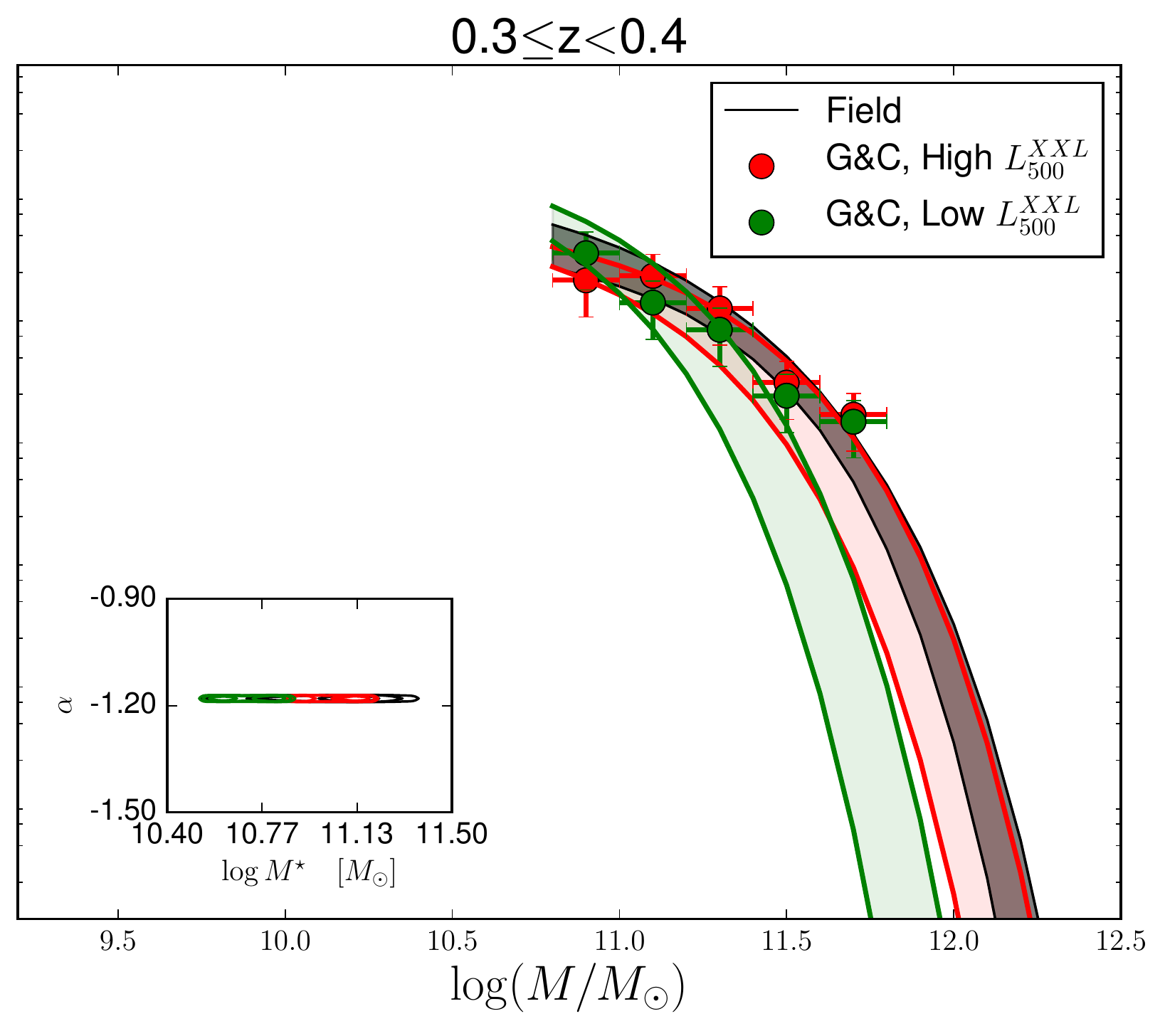}
\caption{Galaxy stellar mass function at different redshifts, as indicated in each panel, for galaxies in G\&C with different X-luminosities and in the field (black diamonds). High X-ray luminosity G\&C members ($L^{XXL}_{500} > 10^{43} {\rm  erg \, s^{-1}}$) are plotted in red; low X-ray luminosity G\&C members are plotted in green. Only points above the mass completeness limit are shown. Error bars on the x-axis show the width of the mass bins; those on the y-axis are derived from Poisson's statistics on the number counts together with the cosmic variance contribution. Due to low number statistics of the sample, we do not show the redshift bin 0.4$\leq z\leq$0.6. At z$\geq$0.3, fixed values for the faint end slope $\alpha$ were set in order to perform Schechter fits.}
\label{mf_Lx}
\end{figure*}

We then build histograms characterising the mass distribution of galaxies located in different environments. For this analysis, in G\&C we use all galaxies within $3r_{200}$.\footnote{The results presented in what follows does not change considerably if we use only galaxies within a distance $r \leq 1.5 \, r_{200}$.} Table \ref{datasample} lists the different samples used.
The width of each mass bin is 0.2 dex. 
In each mass bin, we count the number of galaxies and then we divide this number by the width of the bin, to have the number of galaxies per unit of mass. When building histograms, each galaxy is weighted by its spectroscopic incompleteness correction, as determined in Section \ref{sec_spec_compl}. 
The choice of the mass completeness limit outlined above introduces an additional partial incompleteness in each redshift bin which is redshift dependent. To further correct for this incompleteness, we subdivide each redshift bin into four sub-bins equally spaced in redshift % (e.g. for  $0.1\leq \Delta z \leq 0.2$, we consider  $0.100\leq \Delta z \leq 0.125$,  $0.125\leq \Delta z \leq 0.150$,  $0.150\leq \Delta z \leq 0.175$  and $0.175\leq \Delta z \leq 0.200$) 
and estimate the proper mass completeness limit for each of these sub-bins. We then compute the mass distribution for each of these subsamples separately and assume that the lowest redshift sub-bin %(e.g. $0.100\leq \Delta z \leq 0.125$ among those listed in the previous parenthesis)
does not suffer from incompleteness, and that its mass distribution is thus the real one. The deviations from this shape that were observed in the other three sub-bins must be due to some incompleteness in the mass regime between the adopted and the proper mass limit. We therefore apply a statistical correction forcing the shape of the mass function in each of these sub-bins to be the same as that in the first sub-bin. Specifically, in each sub-bin in redshift, we compute the best-fitting line to the set of counts in the mass range between the adopted mass completeness limit and the proper one. For the first sub-bin, where the proper and adopted mass limit coincide by definition, we perform the fit on the same mass range adopted for the fourth sub-bin, which is the most incomplete. In each sub-bin we then take the ratio of the fit in that sub-bin to the fit in the first sub-bin and we use that factor to correct the number counts in that sub-bin. Finally, the final GSMF in each redshift bin is obtained by summing up all the corrected counts within each mass bin.
We note that this further correction does not introduce any bias in the results that follow. Indeed, performing our analysis considering the original, more conservative, limits we obtain similar results, but with much larger uncertainties.

Galaxy stellar mass functions are normalised using the total integrated stellar mass in the mass range shared by the samples we are comparing, so that the total galaxy stellar mass in each histogram in that mass range is equal to 1. 
This normalization allows us to focus our analysis on the shape of the GSMF and not on the number density, which is obviously very different across the different environments.

In the following plots, error bars on the x-axis represent the width of the bins, error bars along the y-axis are computed adding in quadrature the Poissonian errors \citep{Gehrels1986} and the uncertainties due to cosmic variance, which we compute considering only our field galaxies. Following the procedure explained in \citet{Marchesini2009}, we divided our field into nine subregions and we computed the number density of galaxies of each region separately; the contribution to the error budget from cosmic variance is then $\rm \sigma_{cv} = \phi_i/\sqrt{n}$, where $i$ is any of the stellar mass bins in which the number density is computed and $n$ is the number of sub-regions considered. The uncertainty due to cosmic variance computed using the field sample was also applied to the GSMF in G\&Cs.
Only points above the mass completeness limit are shown. 

First, we test our determination of the GSMF by comparing it with other results from the literature, as shown in Fig. \ref{mf_XXL_Moustakas}. We use as comparison the sample presented in \cite{Moustakas2013}, who exploited multiwavelength imaging and spectroscopic redshifts from the PRism MUlti-object Survey (PRIMUS) over five fields totaling $\sim 5.5$ deg$^2$ to characterise the mass functions in the redshift interval  $0.2<z<1.0$. To increase the statistics, we combine their redshift bins $0.2<z<0.3$ and $0.3<z<0.4$ in the mass range in common between the two and contrast their GSMF to that obtained from the XXL data over the same redshift interval. For this analysis we use both field and G\&C galaxies together, mimicking the analysis of \cite{Moustakas2013}. In this case, the original normalisation over the comoving volume given by \cite{Moustakas2013} was maintained and the values of the GSMF derived in this work were normalised to theirs in the mass range shared by the two curves. Figure \ref{mf_XXL_Moustakas} shows that our GSMF compares remarkably well with the independent determination by \cite{Moustakas2013}, indicating that systematics on the stellar mass determination are under control. We can now proceed with the analysis.

We are now in the position of contrasting the G\&C and field GSMF, as shown in Fig. \ref{mf}, for galaxies at  different redshifts. 
At each cosmic time, the mass distributions in the different environments present a similar shape within the error bars. This result is in agreement with the previous literature data, both in the local universe (e.g. \citealt{Calvi2013}) and at $z\sim0.6$ (e.g. \citealt{Giodini2012,Vulcani2013}).

We note that with increasing redshift and going to higher stellar masses, the GSMF of the field sample changes from being below the G\&C GSMF to being above at the highest redshifts. This trend could be due to the limited statistics of G\&C at higher redshifts and to the detection limit of X-ray observations where we are able to detect only bigger G\&C.

In order to validate our previous statements on the dependence of the GSMF on environment at different redshifts we performed analytical fits to the data points, using a Markov Chain Monte Carlo method. The number density $\rm \Phi (M)$ of galaxies can be described by a Schechter function, given by the equation
\begin{equation}
\Phi(M)=\ln10 \Phi^\star 10^{(M-M^\star)(1+\alpha)}\exp(-10^{(M-M^\star)})
\end{equation}
where $M=\log(M/M_\odot)$, $\alpha$ is the low-mass end slope, $\Phi^\star$ is the normalisation, and $M^\star=\log(M^\star/M_\odot)$ is the characteristic mass. Schechter function fits are computed only above the completeness limits and the best-fit parameters  are reported in Table \ref{Schechter_param}.
A direct hint of the similarity of the GSMF of the samples considered is given by the inset plots included in all panels, which show the confidence contour at 1,2,3 $\sigma$ of the parameters that are significant for our analysis: $\alpha$ and $M^\star$.
At $0.1<z<0.2$, Schechter fits agree within 1$\sigma$, probing on a statistical ground that the shapes of the field and G\&C GSMFs are very similar. Moving to higher redshifts, the significance of the results is lower, but still outstanding  differences do not emerge. Contour levels on the Schechter parameters are superposed at the 2-3$\sigma$ level.
We note that at z$>$0.3, due to the limited mass range probed by our sample, we are not able to probe the slope of the GSMF and hence we can only inspect the exponential tail of the mass distribution. We therefore need to fix the $\alpha$ parameter to reduce the degeneracy and determine $M^*$. We choose the best value that can reproduce our data point distribution, for field and for G\&C galaxies separately.
We caution the reader that comparisons of the parameters while fixing one of the two have to be taken carefully.
Furthermore, at 0.4<z<0.6 we note that $M^\star$ is much less constrained in G\&C than in the field since there are no data points at $\log(M^\star/M_\odot)>12.2$. At lower masses, the two GSMFs clearly overlap.

\begin{figure}
\centering 
\includegraphics[scale=0.45]{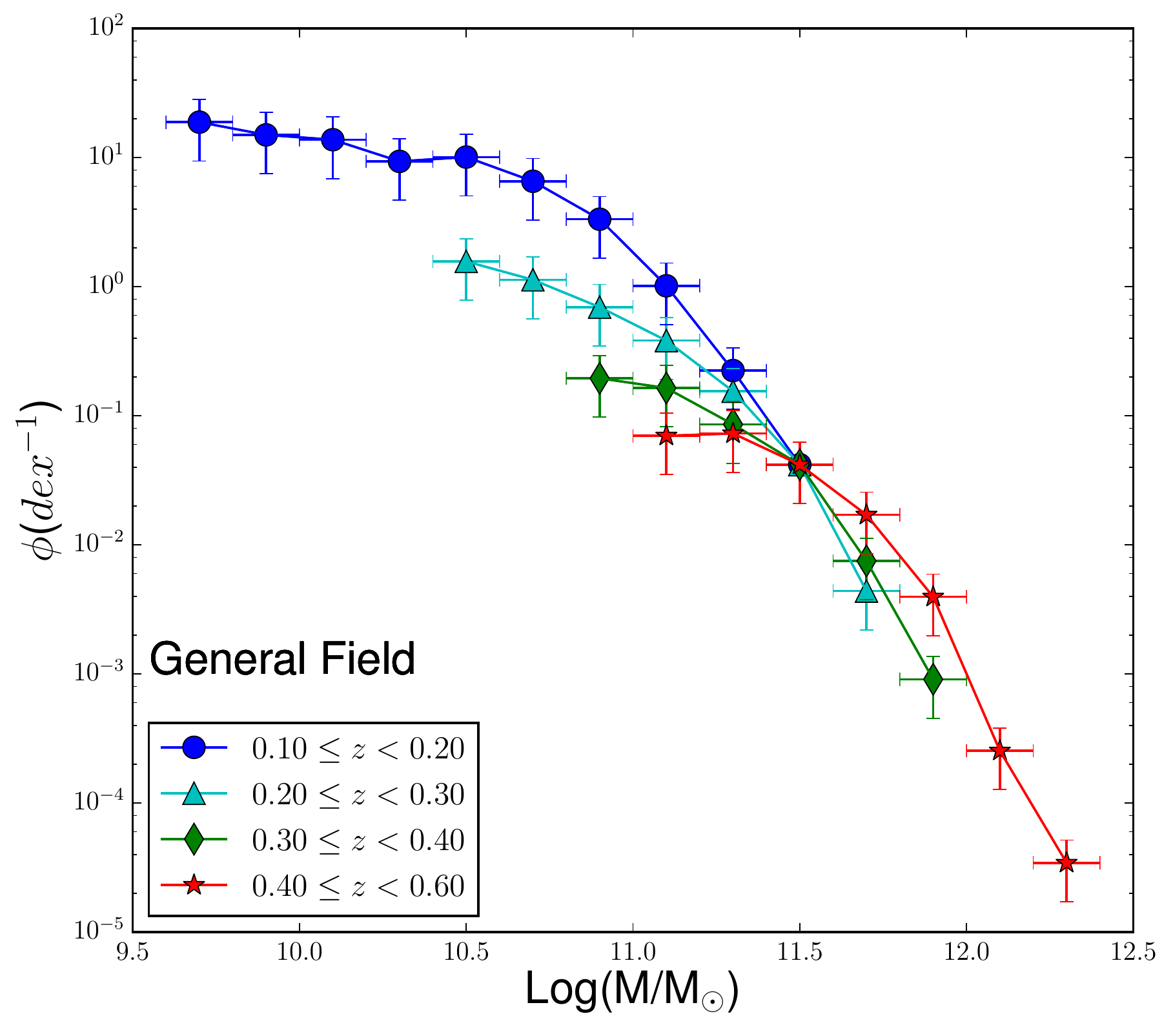}
\caption{Evolution of the GSMF in the general field (pure field$+$G\&C) with redshift. The curves are normalised at the number counts of the highest mass point of the GSMF at $0.1 \leq z < 0.2$ (blue curve).}
\label{mf_z}
\end{figure}

Our G\&C sample spans a wide range of $L^{XXL}_{500}$ (Fig.\ref{mrl500_z}). It is therefore possible to consider separately galaxies in low-luminosity G\&C ($L^{XXL}_{500} <\rm 10^{43} \, {\rm  erg \, s^{-1}}$) and high-luminosity G\&C ($L^{XXL}_{500} >\rm 10^{43} \, {\rm  erg \, s^{-1}}$) and investigate whether the galaxy stellar mass distribution changes with X-ray luminosity. Figure \ref{mf_Lx} shows that  galaxies in G\&C characterised by different values of $L_X$ have very similar mass distribution, emphasising once again how the global environment does not impact the GSMF in the mass range we are probing. These findings are also supported by the analysis of the Schechter fit parameters, shown in the insets of Fig. \ref{mf_Lx} (see also Table \ref{Schechter_param}).
We note that in our sample the number of low-luminosity G\&C at $z>0.4$ is very small; therefore, a statistically meaningful comparison at these redshifts is not possible.
 
Having assessed a similarity in the GSMF for galaxies in the different environments, we can now investigate its evolution with cosmic time. Figure \ref{mf_z} shows the variation of the GSMF with time for the ensemble of the field and G\&C samples. Curves are normalised at the most massive data point at the lowest redshift bin ($0.1 \leq z < 0.2$, blue dots in the figure). In this way we assume that the most massive galaxies are already in place at $z\sim$1 (see e.g. \citealt{Fontana2004}; \citealt{Pozzetti2007}).
%We performed Schechter fits to the data in the first two redshift bins, considering only galaxies in the mass limited sample of the highest bin ($\rm log(M/M_\odot) \geq 10.4$), and we obtained $\alpha$ and $M^*$ that best fit the GSMFs: $\rm \alpha_1=-0.602 \pm 0.001,\, M^*_1=10.8 \pm 0.1 \; and \; \alpha_2=-1.32 \pm 0.04,\, M^*_2=11.04 \pm 0.03$.
Although the mass range sampled at different redshift varies, the GSMFs in the figure show an increase in the relative number of lower mass galaxies with decreasing redshift. 
These results are in agreement with previous findings (e.g. \citealt{Marchesini2009,Moustakas2013,Muzzin2013,Ilbert2013,Vulcani2013}) that showed that while the most massive galaxies are already in place at $z>0.6$, the number of low-mass galaxies proportionally increases going from higher to lower redshift.
We cannot perform Schechter fits on these GSMFs because of the limited number of data points we should rely on. In fact, in order to properly compare the fits, we should consider the stellar mass limit of the highest redshift bin. This condition does not allow both $\alpha$ and $M^*$ parameters to be left free to vary during the fit as we sample only the high-mass end of the GSMF, and would force the assumption of a literature value for the faint end slope of the Schechter function $\alpha$, therefore preventing a direct study on the variation of the number of low-mass galaxies.

The distribution of stellar mass in galaxies in G\&C below $z<1$ was investigated by \cite{Giodini2012}, who exploited 160 X-ray detected galaxy G\&C in the 2 deg$^2$ COSMOS survey at $0.2<z<1$ and determined G\&C memberships with photometric redshifts. Our analysis is  based on a much wider area, reducing the cosmic variance, and on spectroscopic redshifts. 
\cite{Giodini2012} also divided the sample into two subsamples of high- and low-mass G\&C, and in different redshift ranges, probing a wider stellar mass range with respect to our study. Their distribution in X-ray luminosity and virial masses is narrower with respect to our G\&C in the same redshift range, so that we could exploit the dependence of the GSMF on environment also in more massive G\&C. Furthermore, they investigated the shape of the distribution for passive and star-forming galaxies, comparing it to that of the field, and as a consequence a direct comparison with our results cannot be made.
Our studies are therefore complementary.

\section{Correlation between stellar mass and X-ray luminosity}

\begin{figure*}
\begin{center}
\includegraphics[scale=0.4, clip=true,trim=8 5 40 40]{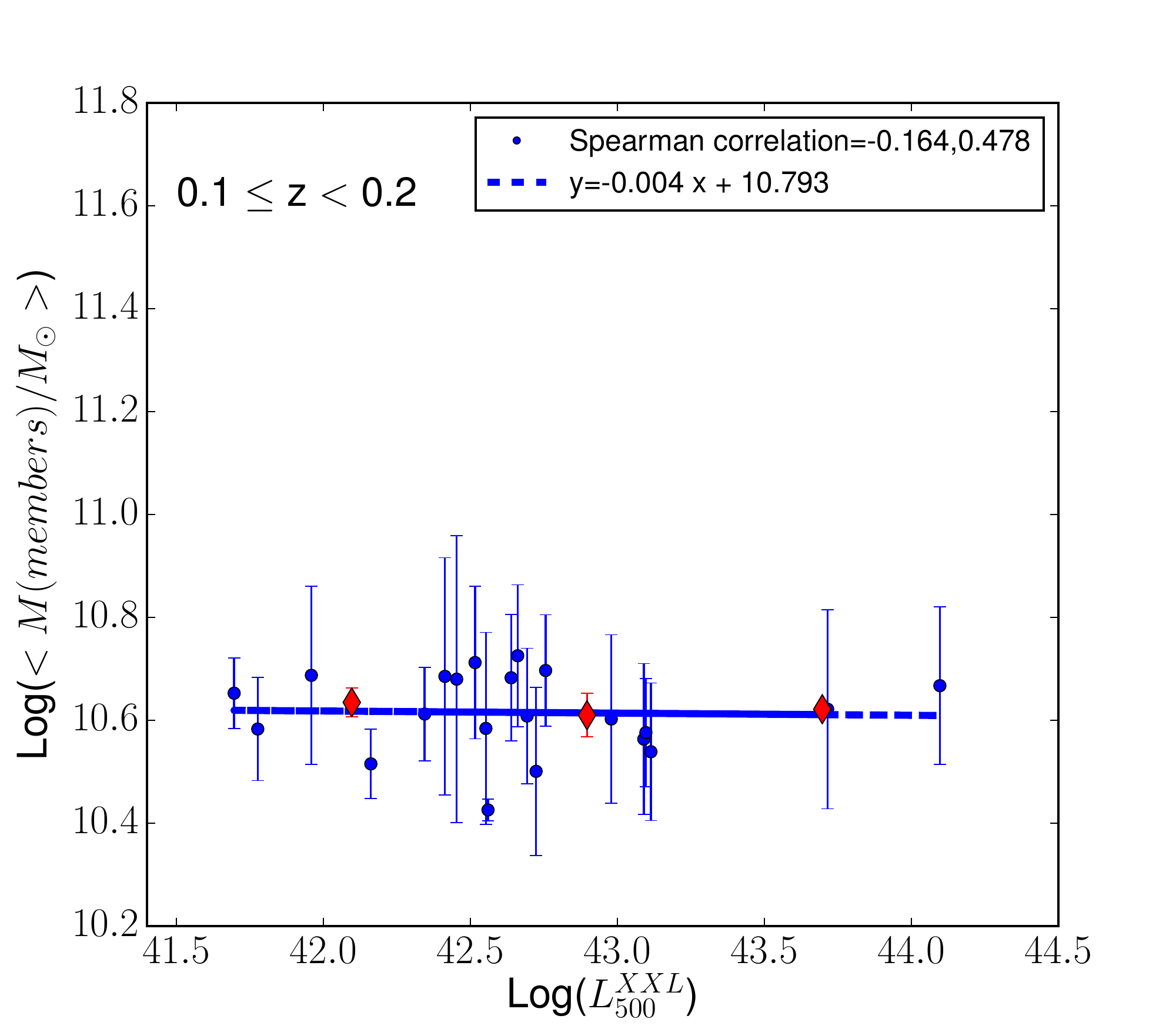}
\includegraphics[scale=0.4, clip=true,trim=8 5 40 40]{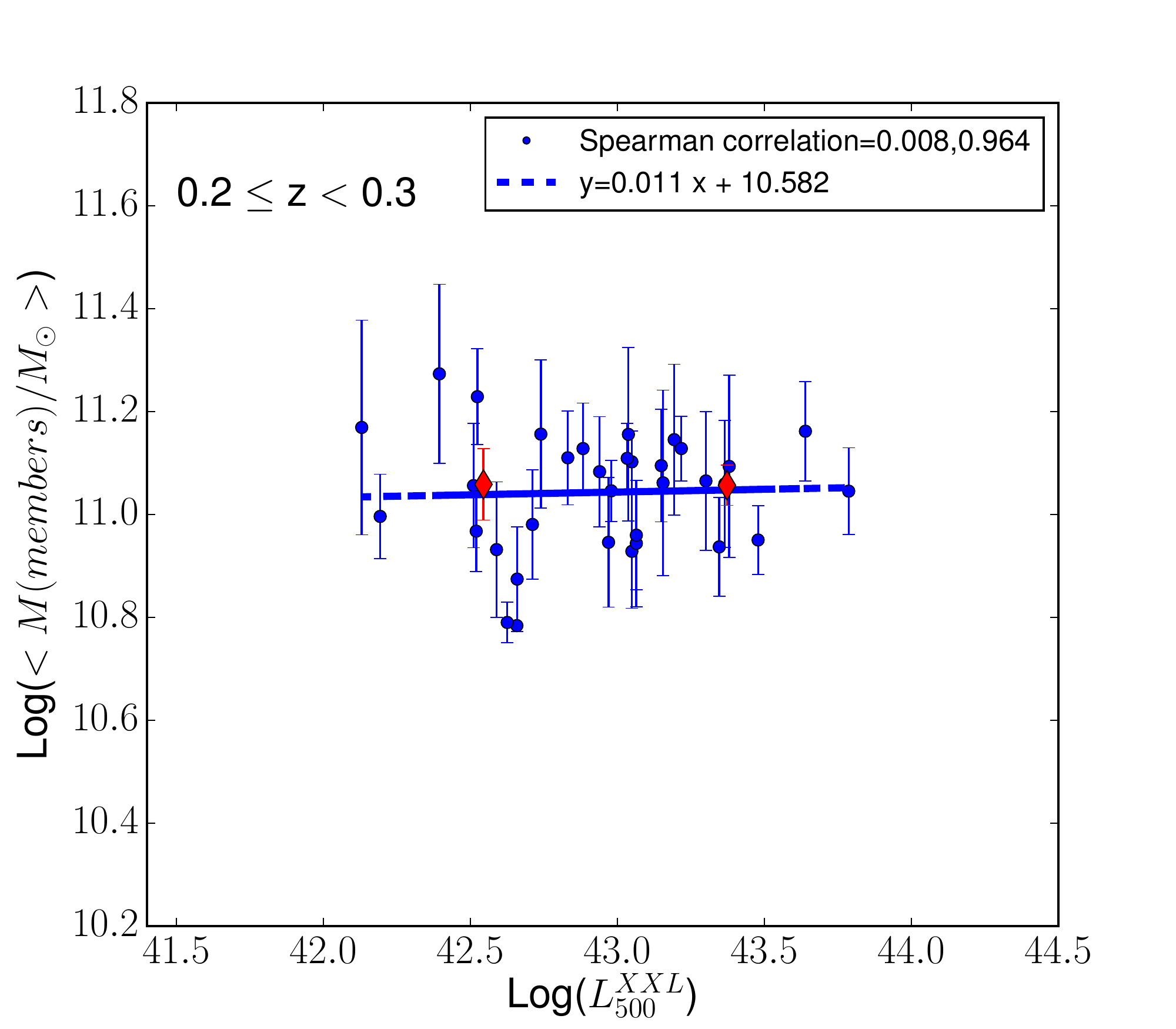}
\includegraphics[scale=0.4, clip=true,trim=8 5 40 40]{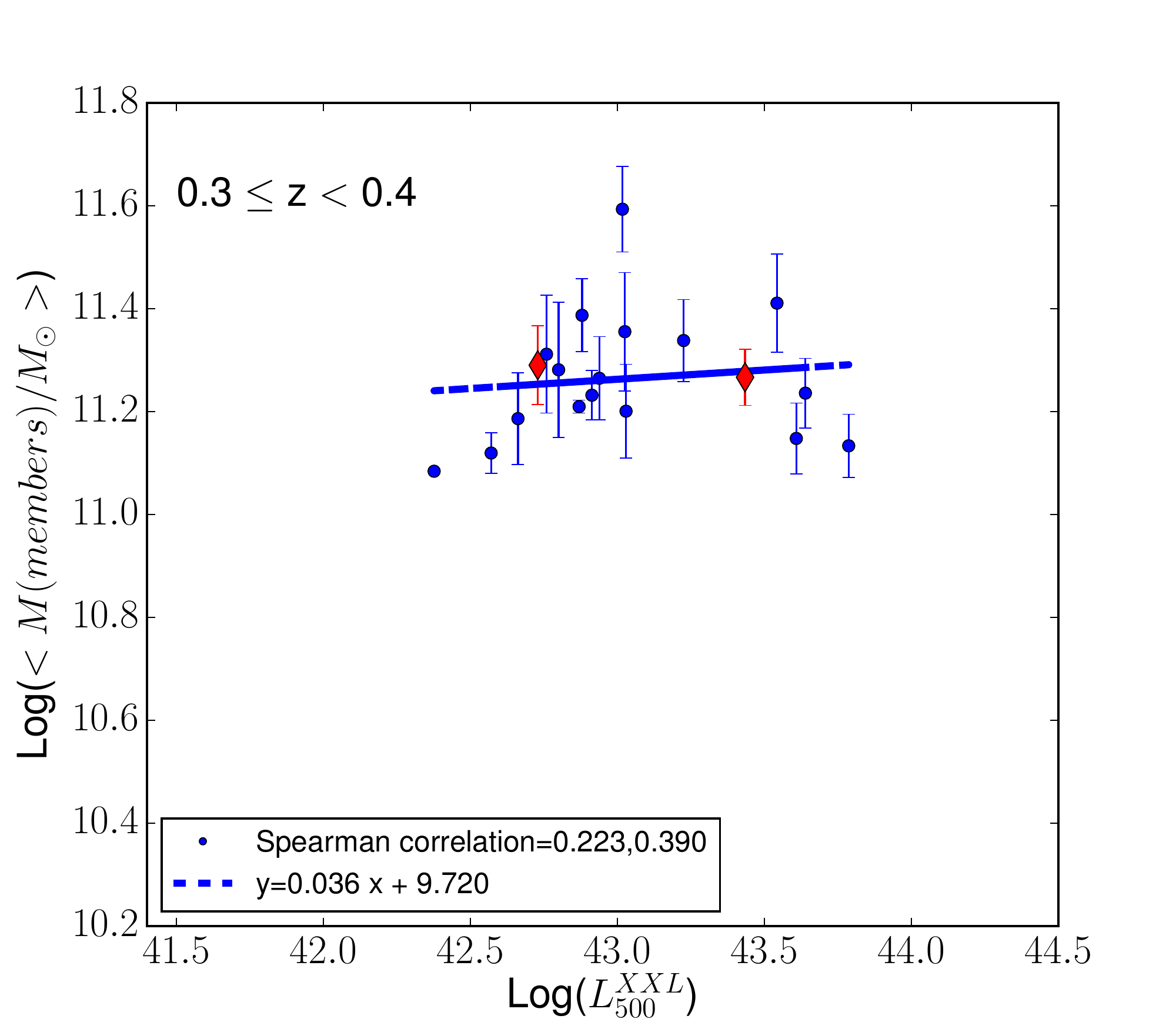}
\includegraphics[scale=0.4, clip=true,trim=8 5 40 40]{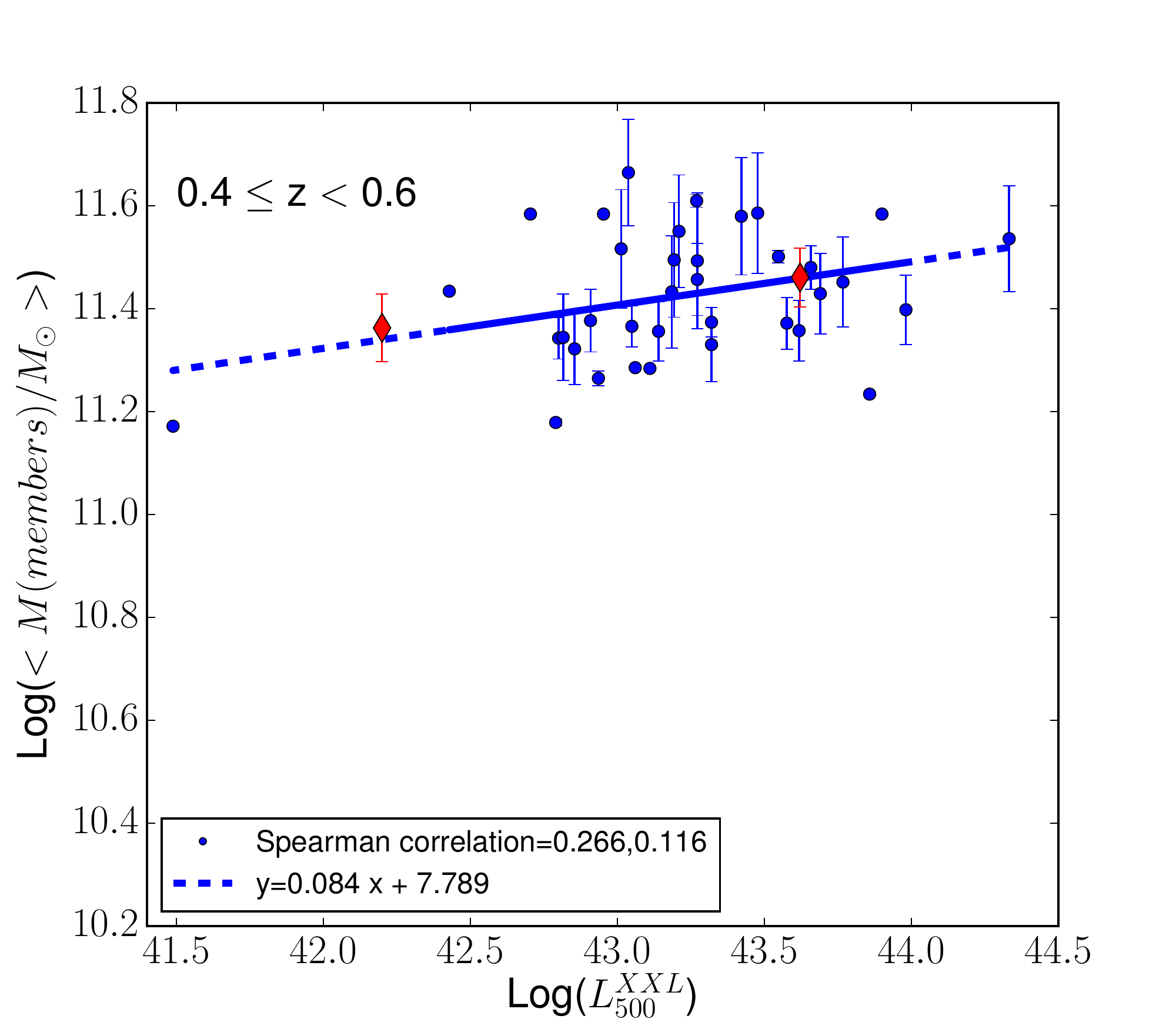}
\caption{Correlation between the mean mass of member galaxies of G\&C and the X-ray luminosity of the host G\&C (blue dots) in the four redshift bins where the stellar mass function was computed. The mean value of the y-axis quantity was computed in equally populated bins of X-ray luminosity (three at $z=0.1-0.2$, two in the other redshift intervals) and is shown with red diamonds. Least-squares fits are shown with dashed lines in the figure and the least-squares fit parameters are shown in the legend.}
\label{m_med_Lx}
\end{center}
\end{figure*}

In the previous section we have shown how the environment has little effect on the overall galaxy stellar mass distribution at least above our mass limit. In addition to the shape of the GSMF, we can also investigate  whether the global properties of the G\&C are related to the typical stellar mass of the galaxies they host. Figure \ref{m_med_Lx} shows the mean stellar mass of G\&C members as a function of the G\&C X-luminosity in the four redshift bins. At each cosmic epoch, mean values are obtained only considering the galaxies that enter the mass complete sample at that redshift. We consider the stellar mass limit of each redshift bin to be the stellar mass limit of the highest redshift subinterval within that bin. We compute the mean value of the mean stellar mass in equally populated bins of X-ray luminosity (three at $z$=0.1-0.2, two in the other redshift intervals). No strong correlations emerge, as also confirmed by the Spearman correlation test.
The first value of the Spearman correlation that is shown in the legend of Figure \ref{m_med_Lx} refers to the slope of the correlation, and the second is the p-value. The latter shows that the correlations are not very tight at all redshifts except the highest one, which is also the only case in which a positive correlation is found. However, we note that the presence of some outliers (e.g. at $z$=0.3-0.4), as well as the scarcity of data in some bins, may influence these results. Least-squares fits of the data are also shown with dashed blue lines in the plot and the least-squares lines are shown in the legend. The slope of the lines in all panels points out that, overall, the trends are almost flat, supporting again the scenario that, at any given redshift, the global environment does not strongly affect galaxy masses.

At similar redshifts, \cite{Vulcani2014}  have shown that in clusters the mass of both the central galaxy and of the most massive satellite correlates with the velocity dispersion of the hosting halo \citep[see also, e.g.,][]{Shankar2006,Wang2006,Moster2010,Leauthaud2010}. They interpreted this evidence as a sign that the environment has a strong effect on the mass of the central and most massive satellites. Indeed, the mass growth of these galaxies is known to be due to mergers and accretion from tidal stripping events, and to different gas cooling and heating mechanisms. All these factors might depend on the size of the G\&C \citep[see, e.g.,][]{Coziol2009, Hopkins2010, Nipoti2012, Newman2012, Vulcani2014b}.

Taken together, these results might indicate that the environment can only affect the mass of peculiar galaxies, like the most massive ones in the systems, but it is not able to impact the overall mass budget. 

Since it is well known that galaxies in different environments and with different stellar masses have different star formation properties and are subject to different physical processes, we should expect different mass growth rates and timescales in different environments. Our findings instead suggest that at the redshifts and mass range considered here, most of the galaxy mass has already been assembled, and that environment-dependent processes have had no significant influence on galaxy mass. This means that at least at $z\leq 0.6$, although strangulation and other gravitational interactions affect other galaxy properties like morphologies and star-forming properties, they have a mild effect on galaxy mass, which has already been assembled, and hence on the galaxy mass distribution. Studies of the properties of the different galaxy populations in the different environments will help in the understanding of the impact of the different processes (Guglielmo et al. in prep.).

\section{Summary}
\label{conclusions}
In this paper we have assembled a catalogue of galaxies in X-ray selected G\&C from the XXL Survey in the redshift range $0<z<1.5$. The XXL Survey is an extension of the XMM-LSS 11 $\rm deg^2$ survey \citep{Pierre2004}, and contains 542 XMM pointings covering a total area of $\sim 50 \, {\rm deg^2}$ reaching a sensitivity of $\sim 5 \times 10^{-15} {\rm erg \, s^{-1} \, cm^{-2}}$ in the [0.5-2] keV band for point sources.

We have mainly focused on the XXL-N region, which covers $\sim$25 deg$^2$. 

The main advantages of our catalogue are the much wider area on the sky compared to other existing catalogues at similar redshift, the X-ray detection, and the spectroscopic confirmation of both the G\&C and of its members, all of which assure robustness. Our G\&C span a wide range of X-ray luminosities ($\rm 2.27 \times 10^{41} \leq L^{XXL}_{500} (erg \, sec^{-1}) \leq 3.18 \times 10^{44}$) and therefore virial masses ($\rm 7.6 \times 10^{12} \leq M_{500} (M_\sun) \leq 6.63 \times 10^{14}$).

Here we have described how both the photometric and spectroscopic samples were assembled and combined. 
We have described the overall properties of the G\&C and the procedure adopted to determine G\&C memberships. We have then computed spectroscopic completeness, stellar masses and stellar mass limits. The catalogue containing the galaxies with $0<z\leq 0.6$ in the magnitude complete sample is made publicly available to the community at CDS and is fully described in Appendix \ref{catalog}.

As a first scientific exploitation of the sample, we have built GSMF for galaxies in G\&C and in the field at different redshifts. As previously found by e.g. \cite{Vulcani2013}, we do not find any significant difference between the shape of the GSMF in the different environments and for galaxies located in G\&C with different X-ray luminosities. 

These findings suggest that at the redshifts considered here environment-dependent processes have had no significant influence on galaxy mass, at least in the mass range we are sampling.

In a future study, we will use the spectrophotometric catalogue presented here to investigate the spectral features of galaxies as a function of redshift and environment, to derive the star formation rate and reconstruct the star formation history within X-ray G\&C, and to compare them with those in the corresponding field sample.

\begin{acknowledgements}
We acknowledge the anonymous referee for the careful report and the important suggestions and comments which helped us to improve our work.
XXL is an international project based around an XMM Very Large Programme surveying two 25 deg$^2$ extragalactic fields at a depth of $\sim 5 \times 10-15 {\rm erg \, s^{-1} \, cm^{-2}}$ in the [0.5-2] keV band for point-like sources. The XXL website is http://irfu.cea.fr/xxl. Multi-band information and spectroscopic follow-up of the X-ray sources were obtained through a number of survey programmes, summarised at http://xxlmultiwave.pbworks.com/. The Australia Telescope Compact Array is part of the Australia Telescope National Facility which is funded by the Australian Government for operation as a National Facility managed by CSIRO.
GAMA is a joint European-Australasian project based around a spectroscopic campaign using the Anglo-Australian Telescope. The GAMA input catalogue is based on data taken from the Sloan Digital Sky Survey and the UKIRT Infrared Deep Sky Survey. Complementary imaging of the GAMA regions is being obtained by a number of independent survey programmes including GALEX MIS, VST KiDS, VISTA VIKING, WISE, Herschel-ATLAS, GMRT and ASKAP providing UV to radio coverage. GAMA is funded by the STFC (UK), the ARC (Australia), the AAO, and the participating institutions. The GAMA website is http://www.gama-survey.org/.
This paper uses data from the VIMOS Public Extragalactic Redshift Survey (VIPERS). VIPERS has been performed using the ESO Very Large Telescope, under the ``Large Programme" 182.A-0886. The participating institutions and funding agencies are listed at http://vipers.inaf.it.

V.G. acknowledges financial support from the Fondazione Ing. Aldo Gini.
B.V. acknowledges the support from an Australian Research Council Discovery Early Career Researcher Award (PD0028506). We acknowledge the financial support from PICS Italy-France scheme (P.I. Angela Iovino).
The Saclay group acknowledges
long-term support from the Centre National d'Etudes Spatiales.
M.E.R.C. and F.P. acknowledge support by the German Aerospace 
Agency (DLR) with funds from the Ministry of Economy and Technology 
(BMWi) through grant 50 OR 1514 and grant 50 OR 1608.
\end{acknowledgements}

\bibliographystyle{aa}
\bibliography{bibliography.bib}

\begin{appendix}

\section{Spectroscopic completeness curves}
\label{app_compl}

\begin{figure}
\begin{center}
\includegraphics[scale=0.28]{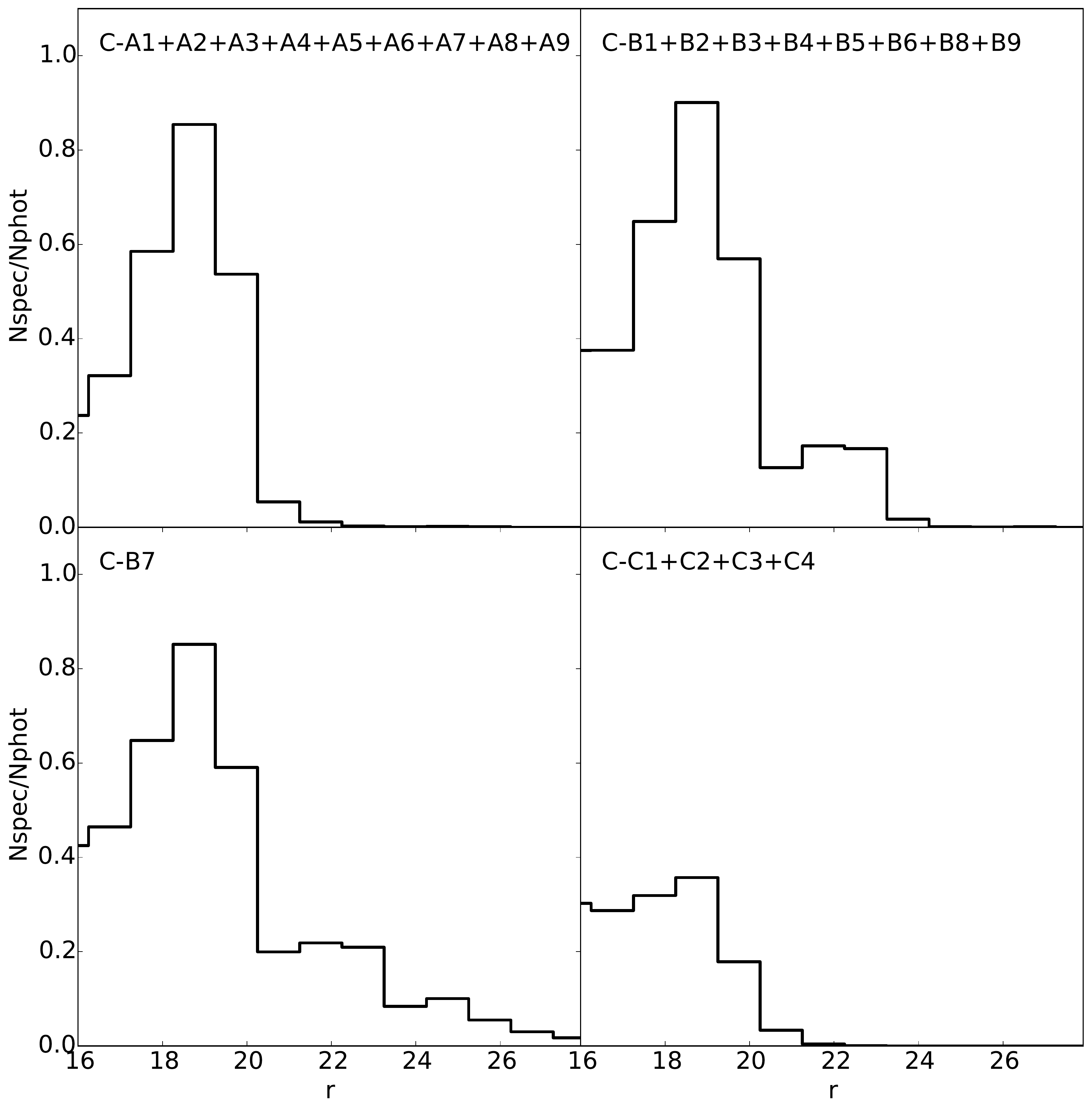}
\caption{Completeness curves as a function of $r$-band magnitude in the four representative regions discussed in the main text, as indicated in each panel.}
\label{compl_curves_all}
\end{center}
\end{figure}

\begin{figure}
\begin{center}
\includegraphics[scale=0.28]{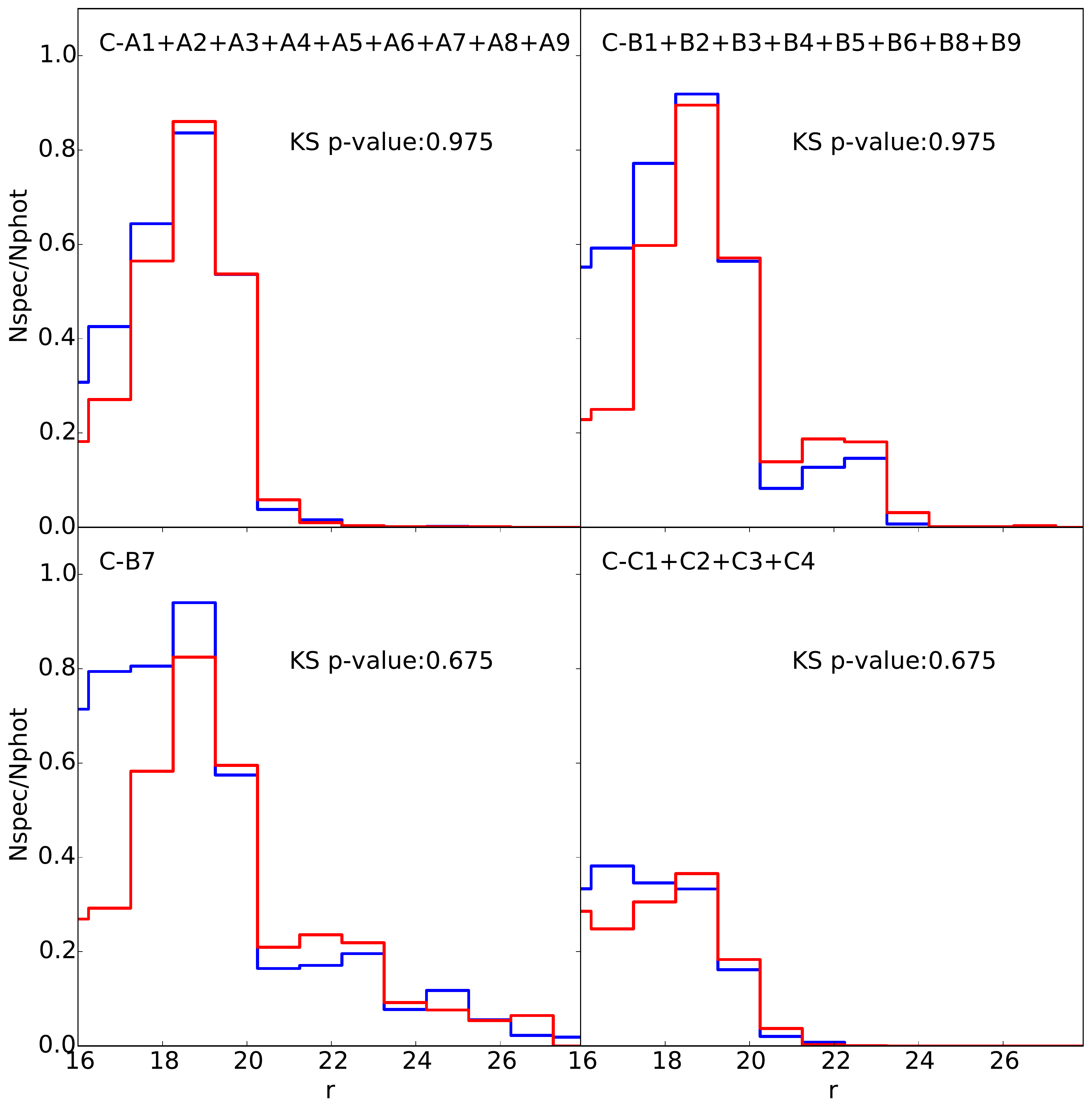}
\caption{Completeness curves as a function of $r$-band magnitude and colour in the four representative regions discussed in the main text, as indicated in each panel. Galaxies are divided into blue and red according to their median observed (g-r) colour. In all the cases, the KS test on the histograms at r$\leq$20 finds no significant differences between the considered samples, as shown by the p-values indicated in each panel.}
\label{compl_curves_col}
\end{center}
\end{figure}

\begin{figure}
\begin{center}
\includegraphics[scale=0.28]{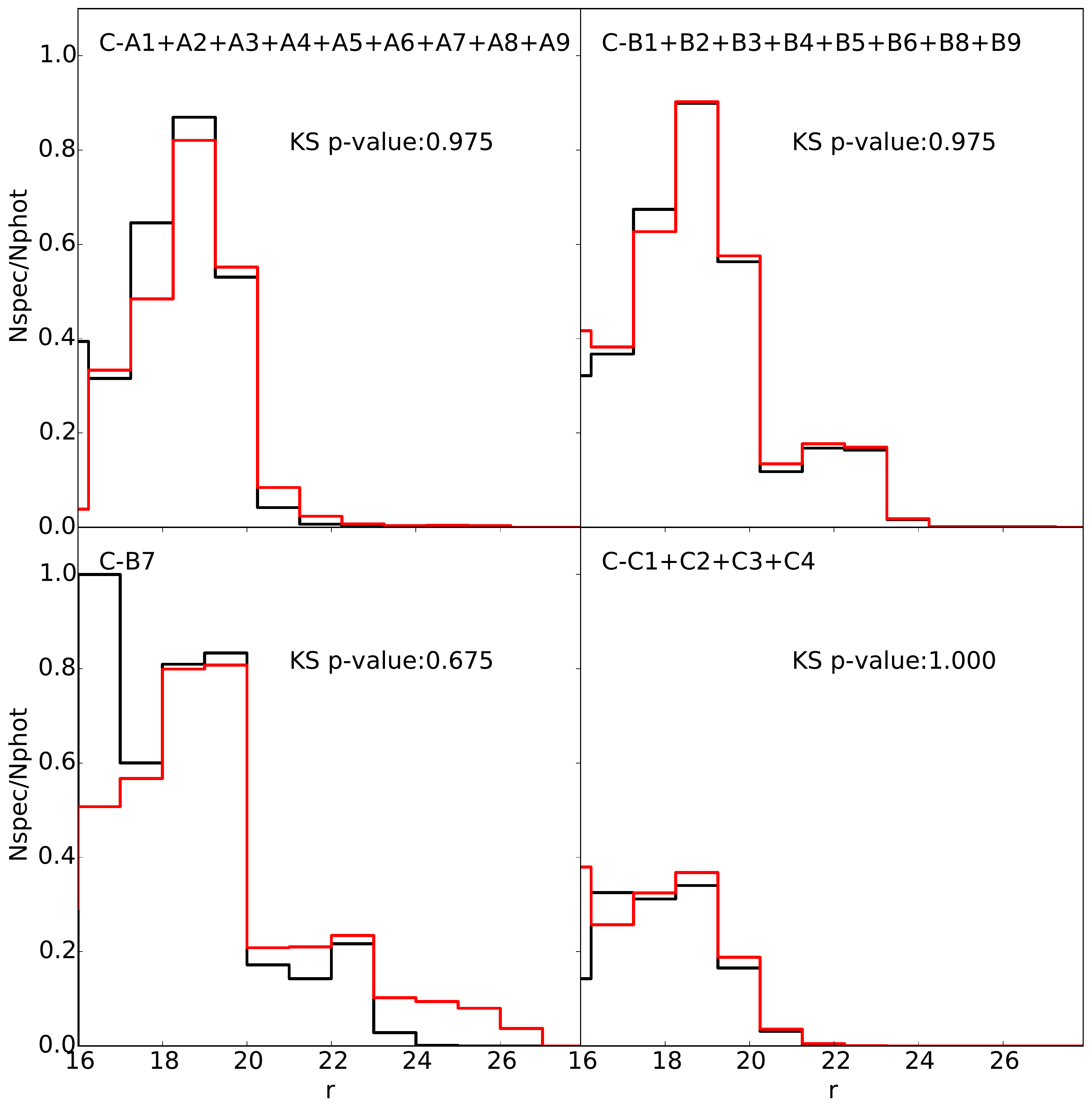}
\caption{Completeness curves as a function of $r$-band magnitude and environment in the four representative regions discussed in the main text, as indicated in each panel. Galaxies in the projected area of G\&C are shown in red, field galaxies are shown in black (see Sect. \ref{membership} for the definitions of the environments). In all the cases, the KS test on the histograms at r$\leq$20 finds no significant differences between the considered samples, as shown by the p-values indicated in each panel.}
\label{compl_curves_env}
\end{center}
\end{figure}

Here we describe in detail the procedure we adopt in Sect. \ref{sec_spec_compl} to compute the spectroscopic completeness of our sample.
As described in the main text, some regions in our survey are not adequately sampled by the available spectroscopy (e.g. the XS regions show lower completeness at any magnitude); therefore, we computed the spectroscopic completeness curves as a function of r-magnitude in each of the 22 cells shown in Fig. \ref{compl_grid}.
Figure \ref{compl_curves_all} shows the curves in four representative regions that gather together contiguous cells showing no differences in their completeness curves. The first region includes the cells in the C-A stripe. The second includes the cells in the C-B stripe, except for the C-B7 cell, where the presence of VVDS and VUDS surveys requires a dedicated analysis. This cell alone constitutes the third region. Finally, the cells in the C-C stripe make up the fourth region. The curves in the four regions highlight how the coverage of the survey is different in different parts of the sky and how our choice of computing the completeness in each cell separately is indeed appropriate.
%\textbf{Figure \ref{compl_curves_all} shows the spectroscopic completeness curves  in four representative regions defined starting from all the cells: C-A1,...,9 cells are summed together to build the first completeness curve, since we did not find relevant differences among the completeness curves. The cells belonging to C-B stripe were also considered all together to build a second representative completeness curve, except for C-B7 cell, where the presence of VVDS and VUDS surveys require a dedicated analysis. Finally, C-C1,2,3,4 cells were summed up to build a completeness curve representative of the last region.}
%We note that  at our magnitudes ($r \leq 20$) the curves are dominated by GAMA spectra
%VIPERS data dominate our sample at faint magnitudes (this survey is responsible to the secondary peak seen at $r\sim22$).
%However, at 

Our adopted magnitude limit ($r= 20$ in the CFHTLS photometry) corresponds to GAMA $r=19.8$, 
%the contribution of VIPERS is negligible 
%(see Section \ref{sec_spec_compl}) 
and GAMA data drive the curves at the magnitudes of interest.

Next, we tested the dependence of the spectroscopic completeness on galaxy colour, drawing completeness ratios as a function of magnitude for blue and red galaxies separately, following the procedure we adopted for the entire sample. We divided the sample into blue and red galaxies, according to the observed (g-r) median colour and computed the spectroscopic completeness for the two populations separately.
We performed a statistical Kolmogorov-Smirnov test (KS) on the resulting completeness curves at r$\leq$20 and found that the two galaxy samples show no significant differences, i.e. the probability that they are drawn from the same parent sample is high, suggesting that our spectroscopic completeness estimates are not biased against any colour.
Figure \ref{compl_curves_col} shows the completeness curves of blue and red galaxies in the sample in the four representative regions. % used at the beginning of the section (Figure \ref{compl_curves_all}). 
The p-values resulting from the KS test on the two samples are shown within each panel.

Finally, we also tested the dependence of the spectroscopic completeness on galaxy environment, to verify whether denser regions in the XXL area have the same sampling as in the field.
We therefore considered separately galaxies in the `pure' field and galaxies that fall into the projected area of G\&C and computed again the spectroscopic weights, following the same method explained in the main text.
A general very good agreement was found between the curves in all regions considered, suggesting that the spectroscopic data almost equally sample regions of different densities in the XXL area, as also supported by the KS test.
However, there are two cases in which the KS test points out a significant difference between the G\&C area and field sample: C-B: $\rm 36.0 < RA \, (deg) \leq 37.0$ (C-B7) and C-A: $\rm 38.0 < RA \,  (deg)\leq 39.0$ (C-A9). This discrepancy can be explained taking into account that the considered areas in the sky are significantly dominated by field and G\&C galaxies, respectively, and therefore the completeness curves of the less populated sample do not have a statistically significant number of objects, either in the photometric or in the spectroscopic sample.
Figure \ref{compl_curves_env} shows the completeness curves of field galaxies and of galaxies in the projected area of G\&C in the four representative regions. %used at the beginning of the section (Figure \ref{compl_curves_all}). 
The p-values resulting from the KS test on the two samples are shown within each panel. As expected, the C-B7 region shows a lower p-value with respect to the other curves; however, it is higher than the commonly adopted p-value used as the threshold that considers the two samples statistically equivalent.

\section{Spectrophotometric catalogue}
\label{catalog}
Here we describe the galaxy catalogue we release, which  contains galaxies in the field and in G\&C at $z\leq 0.6$ with observed magnitude $r\leq 20$. 
The main properties of a subsample of galaxies  are given in Table \ref{cat}, while the total sample can be found at CDS.
The columns indicate the following parameters:
\begin{enumerate}
\item ID: identification sequential number for galaxies.
\item RA: right ascension (deg).
\item DEC: declination (deg).
\item z: redshift from the XXL spectroscopic database.
\item SpecOrigin: parent survey/catalogue of the spectra
\item Origin\_Flag: flag dividing the surveys given in the SpecOrigin column into three classes of priority, as explained in Sect. \ref{spec_data}.
\item Quality\_Flag: flag uniformly dividing zflag values into five classes according to the precision and reliability of the redshift estimate (Section \ref{spec_data}).
\item DeltaR\_r200\_1: distance in units of $\rm r_{200}$ from the first G\&C the galaxy is considered a member of (for field galaxies the value is set to zero).
\item DeltaR\_r200\_2: distance in units of $\rm r_{200}$ from the second G\&C the galaxy is considered a member of (for field galaxies or only single membership the value is set to zero).
\item DeltaR\_r200\_3: distance in units of $\rm r_{200}$ from the third G\&C the galaxy is considered a member of (for field galaxies or only single/double membership the value is set to zero). 
\item DeltaR\_r200\_4: distance in units of $\rm r_{200}$ from the fourth G\&C the galaxy is considered a member of (for field galaxies or only single/double membership the value is set to zero). 
\item DeltaR\_r200\_5: distance in units of $\rm r_{200}$ from the fifth G\&C the galaxy is considered a member of (for field galaxies or only single/double membership the value is set to zero). 
\item Delta\_v\_1: difference in recession velocity from the first G\&C the galaxy is considered a member of (for field galaxies the value is set to zero).
\item Delta\_v\_2: difference in recession velocity from the second G\&C the galaxy is considered a member of (for field galaxies or only single membership the value is set to zero).
\item Delta\_v\_3: difference in recession velocity from the third G\&C the galaxy is considered a member of (for field galaxies or only single/double membership the value is set to zero).
\item Delta\_v\_4: difference in recession velocity from the fourth G\&C the galaxy is considered a member of (for field galaxies or only single/double membership the value is set to zero).
\item Delta\_v\_5: difference in recession velocity from the fifth G\&C the galaxy is considered a member of (for field galaxies or only single/double membership the value is set to zero).
\item XLSSC\_3r200: XLSSC ID of the structure the galaxy belongs to.
In the case of multiple memberships, the multiple identification numbers are separated using the underscore symbol (\_). 
\item XLSSC\_3r200\_uniq: XLSSC ID of the closest G\&C the galaxy belongs to (i.e. the G\&C that minimises the projected distance between the G\&C centre and the galaxy).
\item DeltaR\_r200\_uniq: projected distance in unity of r200 of the closest G\&C given in the previous column. 
\item u\_ABS: rest-frame u-band absolute magnitude computed using LePhare, using spectroscopic redshift and observed magnitudes.
\item g\_ABS: rest-frame g-band absolute magnitude computed using LePhare, using spectroscopic redshift and observed magnitudes.
\item r\_ABS: rest-frame r-band absolute magnitude computed using LePhare, using spectroscopic redshift and observed magnitudes.
\item i\_ABS: rest-frame i-band absolute magnitude computed using LePhare, using spectroscopic redshift and observed magnitudes.
\item y\_ABS: rest-frame y-band absolute magnitude computed using LePhare, using spectroscopic redshift and observed magnitudes.
\item z\_ABS: rest-frame z-band absolute magnitude computed using LePhare, using spectroscopic redshift and observed magnitudes. 
%\item MASS\_BEST: stellar mass at the minimum $\chi^2$ value, as computed from LePhare.
\item MASS\_INF: 16\% lower value on the maximum likelihood (ML) analysis of LePhare
\item MASS\_MED: median value of the stellar mass from the ML analysis of LePhare.
\item MASS\_SUP: 16\% higher value on the ML analysis of LePhare.
\item Compl\_SM: completeness computed using the subsample of the spectrophotometric catalogue including only the galaxies with a reliable stellar mass estimate by LePhare.
\end{enumerate}

In all the columns, we note that the value -99.99 is arbitrarily assigned when the true value is not available.  

\begin{table*}
\begin{center}

\begin{tiny}
\begin{tabular}{lrrrlrr}
\hline
  \multicolumn{1}{c}{ID} &
  \multicolumn{1}{c}{RA} &
  \multicolumn{1}{c}{DEC} &
  \multicolumn{1}{c}{z} &
  \multicolumn{1}{c}{SpecOrigin} &
  \multicolumn{1}{c}{Origin\_Flag} &
  \multicolumn{1}{c}{Quality\_Flag} \\
\hline
  1378 & 31.63026 & -7.56776 & 0.4411 & ESO\_LP & 1 & 400\\
  940 & 34.36603 & -7.70509 & 0.0158 & AAT\_AAOmega\_GAMA & 1 & 400\\
  93017 & 36.10000 & -4.18690 & 0.1065 & VIPERS\_2DR & 1 & 400\\
  1052 & 34.92736 & -7.66780 & 0.1082 & SDSS\_DR10 & 1 & 400\\
  59658 & 35.50422 & -4.80558 & 0.2050 & AAT\_AAOmega\_GAMA & 1 & 400\\
  100987 & 30.54604 & -4.99444 & 0.2340 & AAT\_AAOmega\_GAMA & 1 & 400\\
  99479 & 37.66412 & -4.96348 & 0.2867 & AAT\_AAOmega\_GAMA & 1 & 400\\
  99777 & 37.65939 & -4.95309 & 0.2898 & AAT\_AAOmega\_GAMA & 1 & 400\\
  99540 & 32.72662 & -6.22625 & 0.4218 & AAT\_AAOmega\_GAMA & 1 & 400\\
  98614 & 32.80906 & -6.15934 & 0.4235 & WHT & 1 & 2\\
\hline
\vspace{2.5mm}
\end{tabular}

\begin{tabular}{lrrrrr}
\hline
  \multicolumn{1}{c}{ID} &
  \multicolumn{1}{c}{DeltaR\_R200\_1} &
  \multicolumn{1}{c}{DeltaR\_R200\_2} &
  \multicolumn{1}{c}{DeltaR\_R200\_3} &
  \multicolumn{1}{c}{DeltaR\_R200\_4} &
  \multicolumn{1}{c}{DeltaR\_R200\_5} \\
\hline
  1378 & 0.0 & 0.0 & 0.0 & 0.0 & 0.0\\
  940 & 0.0 & 0.0 & 0.0 & 0.0 & 0.0\\
  93017 & 0.0 & 0.0 & 0.0 & 0.0 & 0.0\\
  1052 & 0.0 & 0.0 & 0.0 & 0.0 & 0.0\\
  59658 & 0.0 & 0.0 & 0.0 & 0.0 & 0.0\\
  100987 & 1.67932 & 0.0 & 0.0 & 0.0 & 0.0\\
  99479 & 0.67039 & 0.47195 & 0.0 & 0.0 & 0.0\\
  99777 & 2.04885 & 0.74780 & 0.64046 & 0.0 & 0.0\\
  99540 & 1.17331 & 0.50395 & 2.61695 & 2.40155 & 0.0\\
  98614 & 2.04803 & 1.53839 & 1.35872 & 1.27156 & 0.06218\\
\hline
\vspace{2.5mm}
\end{tabular}

\begin{tabular}{lrrrrrl}
\hline
  \multicolumn{1}{c}{ID} &
  \multicolumn{1}{c}{Delta\_v\_1} &
  \multicolumn{1}{c}{Delta\_v\_2} &
  \multicolumn{1}{c}{Delta\_v\_3} &
  \multicolumn{1}{c}{Delta\_v\_4} &
  \multicolumn{1}{c}{Delta\_v\_5} &
  \multicolumn{1}{c}{XLSSC\_3r200} \\
\hline
  1378 & 0.0 & 0.0 & 0.0 & 0.0 & 0.0 & 0\\
  940 & 0.0 & 0.0 & 0.0 & 0.0 & 0.0 & 0\\
  93017 & 0.0 & 0.0 & 0.0 & 0.0 & 0.0 & 0\\
  1052 & 0.0 & 0.0 & 0.0 & 0.0 & 0.0 & 0\\
  59658 & 0.0 & 0.0 & 0.0 & 0.0 & 0.0 & 0\\
  100987 & 116.6602 & 0.0 & 0.0 & 0.0 & 0.0 & 114\\
  99479 & 1181.2538 & 1181.2538 & 0.0 & 0.0 & 0.0 & 149\_150\\
  99777 & 926.8587 & 464.1469 & 464.1469 & 0.0 & 0.0 & 148\_149\_150\\
  99540 & 1050.5763 & 1698.2439 & 1259.8086 & 505.1972 & 0.0 & 082\_083\_085\_086\\
  98614 & 693.3804 & 1341.8223 & 1279.1928 & 902.8628 & 147.3492 & 082\_083\_084\_085\_086\\
\hline
\vspace{2.5mm}
\end{tabular}

\begin{tabular}{lrrrrrrrr}
\hline
  \multicolumn{1}{c}{ID} &
  \multicolumn{1}{c}{XLSSC\_3r200\_uniq} &
  \multicolumn{1}{c}{DeltaR\_r200\_uniq} &
  \multicolumn{1}{c}{u\_ABS} &
  \multicolumn{1}{c}{g\_ABS} &
  \multicolumn{1}{c}{r\_ABS} &
  \multicolumn{1}{c}{i\_ABS} &
  \multicolumn{1}{c}{y\_ABS} &
  \multicolumn{1}{c}{z\_ABS} \\
\hline
  1378 & 0 & 0.0 & -20.338 & -20.786 & -21.096 & -21.318 & -21.298 & -21.499\\
  940 & 0 & 0.0 & -15.621 & -16.186 & -16.501 & -16.697 & -16.681 & -16.745\\
  93017 & 0 & 0.0 & -15.513 & -15.680 & -16.458 & -16.596 & -16.572 & -16.811\\
  1052 & 0 & 0.0 & -18.768 & -20.115 & -20.833 & -21.245 & -21.208 & -21.552\\
  59658 & 0 & 0.0 & -19.542 & -20.130 & -20.461 & -20.632 & -20.614 & -20.802\\
  100987 & 114 & 1.67932 & -19.357 & -20.204 & -20.811 & -21.150 & -21.118 & -21.415\\
  99479 & 150 & 0.47195 & -20.448 & -21.184 & -21.647 & -22.024 & -21.996 & -22.264\\
  99777 & 150 & 0.64046 & -20.217 & -21.439 & -22.112 & -22.513 & -22.481 & -22.790\\
  99540 & 083 & 0.50395 & -20.558 & -21.734 & -22.344 & -22.658 & -22.631 & -22.893\\
  98614 & 086 & 0.06218 & -20.873 & -22.149 & -22.812 & -23.126 & -23.098 & -23.360\\
\hline
\vspace{2.5mm}
\end{tabular}

\begin{tabular}{lrrrrr}
\hline
  \multicolumn{1}{c}{ID} &
  \multicolumn{1}{c}{MASS\_INF} &
  \multicolumn{1}{c}{MASS\_MED} &
  \multicolumn{1}{c}{MASS\_SUP} &
  \multicolumn{1}{c}{Compl\_SM} \\
\hline
  1378 & 9.708 & 9.747 & 9.789 & 0.0346\\
  940 & 8.502 & 8.536 & 8.578 &  0.0\\
  93017 & 8.520 & 8.633 & 8.708 & 0.2331\\
  1052 & 10.516 & 10.550 & 10.584 & 0.0\\
  59658 & 9.569 & 9.608 & 9.642 &  0.7011\\
  100987 & 10.119 & 10.174 & 10.254 &  0.6941\\
  99479 & 10.167 & 10.202 & 10.236 &  0.8812\\
  99777 & 11.016 & 11.050 & 11.084 &  0.8812\\
  99540 & 11.016 & 11.050 & 11.084 &  0.0819\\
  98614 & 11.166 & 11.200 & 11.234 &  0.2412\\
\hline
\vspace{2.5mm}
\end{tabular}

\end{tiny}
\caption{Subsample of ten galaxies in the catalogue with their properties. The full table can be found at CDS. The explanation of the different columns is given in Appendix \ref{catalog}. The column `ID' is repeated at the beginning of each part of the table for the sake of clarity.  
\label{cat}}
\end{center}
\end{table*}

\end{appendix}

\end{document}